\documentclass[aps,prb,floatfix,twocolumn]{revtex4-2}
\usepackage{graphicx}
\usepackage{bm,amsmath}
\usepackage{dcolumn}
\usepackage{amsfonts,amssymb}
\usepackage{chemfig}

\begin{document}

\title{Dynamic structure factor of a spin-1/2 Heisenberg chain with long-range interactions}
\author{Sibin Yang}
\affiliation{Department of Physics, Boston University, 590 Commonwealth Avenue, Boston, Massachusetts 02215, USA}

\author{Gabe Schumm}
\affiliation{Department of Physics, Boston University, 590 Commonwealth Avenue, Boston, Massachusetts 02215, USA}

\author{Anders W. Sandvik}
\email{sandvik@buphy.bu.edu}
\affiliation{Department of Physics, Boston University, 590 Commonwealth Avenue, Boston, Massachusetts 02215, USA}

\date{\today}

\begin{abstract}
  We study the dynamic structure factor $S(k,\omega)$ of the spin-1/2 chain with long-range, power-law decaying unfrustrated (sign
  alternating) Heisenberg interactions $J_r \sim (-1)^{r-1} r^{-\alpha}$ by means of stochastic analytic continuation (SAC) of imaginary-time
  correlations computed by quantum Monte Carlo calculations. We do so in both the long-range antiferromagnetic (AFM, for $\alpha \alt 2.23$) and
  quasi-long-range-ordered (QLRO, for $\alpha \agt 2.23$) ground-state phases, employing different SAC parametrizations of $S(k,\omega)$ to resolve sharp
  edges characteristic of fractional quasi-particles and sharp peaks expected with conventional quasi-particles. In order to identify the most
  statistically accurate parametrization, we apply a newly developed cross-validation method as a ``model selection'' tool. We confirm that the
  spectral function contains a power-law divergent edge in the QLRO phase and a very sharp (likely $\delta$-function) magnon peak in the AFM phase.
  From our SAC results, we extract the dispersion relation in the different regimes of the model, and in the AFM phase we extract the weight of
  the magnon pole. In the limit where the model reduces to the conventional Heisenberg chain with nearest-neighbor interactions, our $S(k,\omega)$
  agrees well with known Bethe ansatz results. In the AFM phase the low-energy dispersion relation is known to be nonlinear, $\omega_k \sim k^z$,
  and we extract the corresponding dynamic exponent $z(\alpha)$, which in general is somewhat above the form obtained in linear spin-wave
  theory. We also find a significant continuum above the magnon peak. This study serves as a benchmark for SAC/QMC studies of systems with a transition
  from conventional to fractionalized quasi-particles.
\end{abstract}
\maketitle

\section{Introduction}\label{sec:intro}

With the development of experimental techniques to accurately manipulate quantum many-body systems, a growing interest in low dimensional
systems with long-range interactions has emerged. Experiments utilizing trapped ions, Rydberg atoms, and quantum gases
\cite{Blatt12,Britton12,Baumann10,Barredo18,Endres16,Labuhn16,Feng23,Chen23}
have stimulated new theoretical exploration of related fundamental model Hamiltonians. One such model that has garnered interest due to its direct
experimental realization is the spin-1/2 chain with power-law decaying, long-range interactions of the form $\sim 1/r^\alpha$. Feng
\textit{et al.}~\cite{Feng23} have successfully prepared quantum XY spin chains with this form of the interactions in trapped-ion systems
and observed continuous symmetry breaking, which would not be possible with short-range interactions according to the Mermin-Wagner
theorem \cite{Mermin66}. It was claimed that the experimental platform can be extended to also study the XXZ model, where
some theoretical work has already been done \cite{Maghrebi17}. 

Another potential realization of long-range interactions in one dimension is transition metal atomic chains \cite{Tung11},
for which an $\it{ab}$ $\it{initio}$ study found long-range decaying Heisenberg (XXX) exchange interactions with alternating (unfrustrated,
bipartite) signs $(-1)^r$. With the distance- ($r-$) dependent interactions of the form $J_r \sim (-1)^{r-1} r^{-\alpha}$, the phase diagram of the
the model with $S=1/2$ spins, which we will focus on here, contains a long-range ordered N\'eel antiferromagnetic (AFM) phase in addition
to the quasi-long-range ordered (QLRO) critical phase exemplified by the conventional Heisenberg chain corresponding to $\alpha \to \infty$
settled \cite{Yusuf04,Laflorencie05,Sandvik10,Zhao24,Adelhardt20,Adelhardt23,Adelhardt24}.

With the alternating signs of the interaction versus $r$, there is no frustration and, therefore, the model is amenable to large-scale quantum
Monte Carlo (QMC) calculations. We study the Hamiltonian in the form
\begin{equation}\label{hamiltonian}
 H=\sum_{r=1}^{L/2}J_r\sum_{i=1}^{L}\textbf{S}_i\cdot\textbf{S}_{i+r},
\end{equation}
where the distance dependent coupling $J_r$ governed by the exponent $\alpha$ is given by
\begin{equation}\label{jrdef}
 J_r=G\frac{(-1)^{r-1}}{r^\alpha}, \quad G=\left (1+\sum_{r=2}^{L/2}\frac{1}{r^\alpha} \right )^{-1}.
\end{equation}
Here $G$ is a normalization factor chosen such that the summation of $|J_r|$ is equal to unity. Without this normalization (i.e., with $G=1$)
the energy is superextensive for $\alpha \le 1$ and our definition of $G$ makes it more practical to study the model for any $\alpha \ge  0$.

\begin{figure}[t]
\includegraphics[width=8cm]{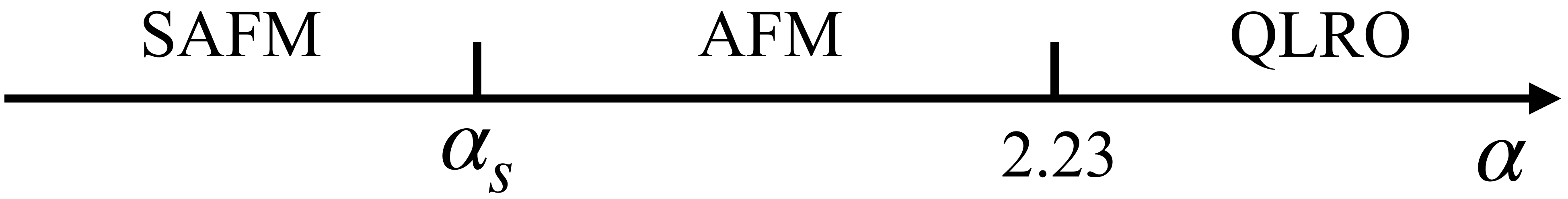}
\caption{Phase diagram of the spin-1/2 Heisenberg chain with long-range interactions [Eq.~(\ref{hamiltonian})]. The critical value of the long-range exponent
$\alpha$ separating the AFM and QLRO phases is $\alpha_c\approx 2.23$ and there is another point $\alpha_s < \alpha_c$ separating the AFM and SAFM phases.
In linear spin wave theory $\alpha_s=1$ \cite{Yusuf04,Laflorencie05} but the exact value may be different.}
\label{Fig.phasediagram}
\end{figure} 

Previous works have identified three ground state phases of the model \cite{Frohlich78,Parreira97,Yusuf04,Laflorencie05}.
For large $\alpha$, the long-range interaction is irrelevant and the ground state is QLRO with the same critical behavior as the standard
Heisenberg chain with only $J_1$ interactions. Upon lowering $\alpha$, there is a quantum phase transition at $\alpha=\alpha_c \approx 2.23$
\cite{Laflorencie05,Zhao24} to a long-range ordered AFM phase, in which the order parameter,
the staggered magnetization $m_s$, increases as $\alpha$ is further reduced until saturation at $m_s=1/2$, occurring when $\alpha = \alpha_s > 0$. Based on spin-wave theory (SWT), the saturation point is exactly at $\alpha_s=1$ \cite{Yusuf04,Laflorencie05}. We will refer to the phase
for $\alpha \in [0,1]$ as the saturated AFM (SAFM) phase. A simple phase diagram summarizing this information is shown in Fig.~\ref{Fig.phasediagram}.

Our focus here is on the dynamic spin structure factor $S(k,\omega)$, which we compute using stochastic analytic continuation (SAC) of
imaginary-time dependent correlation functions generated by the stochastic series expansion (SSE) QMC method. Though the ground state
phase diagram of the model is now settled and unusual properties of the AFM phase and the QLRO--AFM transition phase have been investigated
extensively \cite{Laflorencie05,Sandvik10,Zhao24}, dynamical properties have been
studied in less detail. In the AFM phase, linear SWT predicts that $S(k,\omega)$ contains a single $\delta$-peak at
energy $\omega_k$, with an anomalous dispersion relation $\omega_k \propto k^z$ and $\omega_{\pi-q} \propto q^z$ in the neighborhood of the two gapless
points $k=0,\pi$. Here, the dynamic exponent $z < 1$, in contrast to $z=1$ for conventional spin waves in higher dimensions.

The only previous numerical study of the dynamic structure factor that we are aware of is the recent time dependent density-matrix
renormalization group (tDMRG) work by Yang et al.~\cite{Yang21}, who investigated the evolution of the deconfined spinons
of the QLRO phase into coherent magnons in the AFM phase. The results demonstrated the expected salient spectral features of a dominant
magnon peak in the AFM phase and more continuum spectral weight in the QLRO phase. However, strong finite-size effects and broadening from
the limitations on the tDMRG time evolution prohibited a more precise determination of the full spectral profile and the anomalous magnon
dispersion in the AFM phase. Here we study $S(k,\omega)$ on larger periodic chains (instead of the open boundaries in the tDMRG study), using
QMC simulations to compute the corresponding spin correlation function in imaginary time, which we subsequently continue to real frequency
using the SAC method \cite{Sandvik16,Shao17,Shao23}.

Our goal with this work is two-fold: to obtain reliable extensive size converged results in the AFM and QLRO ground states of the model
and to demonstrate the capabilities of newly developed enhancements of the SAC method \cite{Shao23,Schumm24}. We adopt several different parametrizations of
the sampled $S(k,\omega)$, with constraints based on prior knowledge or hypotheses of the two ground state phases. The ability of constrained
SAC to reproduce detailed sharp features of spectral functions \cite{Shao23} is important not only for comparison with future experiments
addressing the excitation spectrum, specifically inelastic neutron scattering, but will also provide guidance to resolving long-standing
theoretical uncertainty about the form of the spectral peak in the AFM and QLRO phase.

The main question here is whether the spectral
function in the AFM phase is dominated by a sharp quasi-particle peak or by an edge followed by a continuum, the latter being a well
known consequence of deconfined spinons in the QLRO phase but could also potentially be realized in some form with anomalous magnons.
Inspired by ideas exploited in machine learning, we have developed a cross-validation method within the SAC framework, which is able to determine the best possible functional form from different SAC results. This method was described in detail in Ref.~\onlinecite{Schumm24},
where we also presented some preliminary tests on the model considered here. We here present a much more comprehensive range of calculations
and focus on the physics of the model.

The rest of the paper is organized as follows: In Sec.~\ref{sec:sac} we introduce the SAC method used to extract the spectral function
using QMC-computed imaginary-time correlation data $G(k,\tau)$. We extract $S(k,\omega)$ in the AFM and SAFM phases in Sec.~\ref{sec:SpectrumAFM}
and in the QLRO phase in Sec.~\ref{sec:sacpowerlaw}, comparing results obtained with different SAC parametrizations. The cross-validation
method is briefly introduced in Sec.~\ref{sec:cross_val}, followed by the use of the method to determine the best parametrization and our
final conclusion for the form of $S(k,\omega)$ in the AFM and QLRO phases. The non-universal dynamic exponent of the dispersion relation
$\omega_k \sim k^z$ in the AFM phase is extracted versus $\alpha$ and compared with linear spin-wave results in Sec.~\ref{sec:dispersionAFM},
with close attention paid to the dependence on the system size. We summarize and further discuss the results and their implications in
Sec.~\ref{sec:summary}. Some auxiliary calculations are reported in three appendixes.

\section{Stochastic Analytic Continuation}
\label{sec:sac}

A general spectral function of an operator $\hat{O}$ can be expressed using the eigenstates $|n\rangle$ and energies $E_n$ as
\begin{equation} {\label{eq.strucfactor}}
 S(\omega)=\frac{\pi}{Z}\sum_{m,n}e^{-\beta E_n} |\langle m|\hat{O}|n\rangle|^2\delta(\omega-[E_m-E_n]),
\end{equation}
where $Z$ is the partition function and $\beta = T^{-1}$ is the inverse temperature.

For a bosonic operator $\hat{O}$ of interest, the imaginary-time correlation function computable by QMC simulations
is defined as
\begin{equation}
 G(\tau)=\langle \hat{O}^\dag (\tau) \hat{O}(0) \rangle,
\end{equation}
where the time evolved operator is given by
\begin{equation}
\hat{O}(\tau)=\text{e}^{\tau H}\hat{O}\text{e}^{-\tau H},
\end{equation}
setting $\hbar=1$. The formal relationship between $G(\tau)$ and the real-frequency spectral function $S(\omega)$ is
\begin{equation}\label{eq.Gtau}
G(\tau)=\frac{1}{\pi}\int_{-\infty}^{+\infty} d\omega S(\omega) e^{-\tau \omega},
\end{equation}
where $\tau \in [0,\beta]$ and $G(\tau)=G(\beta-\tau)$.

Here we consider the dynamic spin structure factor $S(k,\omega)$ of the spin chain, for which $\hat{O}$ is the Fourier
transform of the $z$-component of the spin operator;
\begin{equation}
\hat{O}=\hat{S}_{\textbf{k}}^z=\frac{1}{\sqrt{N}}\sum_{i=1}^{N}e^{-i\textbf{r}_i\cdot\textbf{k}}\hat{S}_i^z.
\end{equation}
For a bosonic operator $\hat{O}$ as the above, the spectral weights at positive and negative energies are related according to $S(-\omega)=e^{-\beta \omega}S(\omega)$.
In practice, we then use an alternative form to Eq.~(\ref{eq.Gtau});
\begin{equation}\label{eq.Gtau2}
G(\tau)=\frac{1}{\pi}\int_{0}^{\infty} d\omega S(\omega) (e^{-\tau \omega}+e^{-(\beta-\tau) \omega}).
\end{equation}
Since we will work effectively in the $T \to 0$ limit, the kernel is still very close to just $e^{-\tau \omega}$ and
the spectral weight at $\omega < 0$ is vanishingly small. We can then normalize the total spectral weight of $S(\omega)$ to unity by dividing
the correlation functions $G(\tau_i)$ at all time points $\tau_i$ considered by $G(0)$. The reason to do this is simply for convenience when working
with $S(\omega)$ in the program; the original normalization $G(0)$ is multiplied back in for the final $S(\omega)$. See Ref.~\cite{Shao23} for slightly
different procedures at nonzero temperatures.

\begin{figure}[t]
\includegraphics[width=8.3cm]{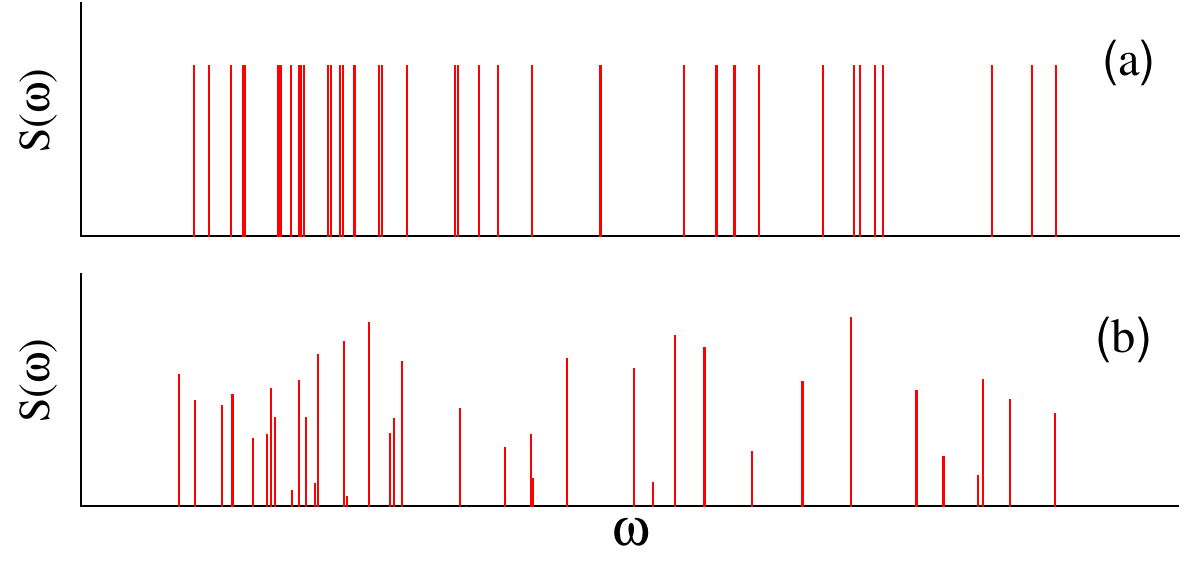}
\caption{Two parametrizations of the spectral function $S(\omega)$ used in unconstrained SAC. In (a), the locations $\omega_i$ of
equal-amplitude $\delta$-functions are sampled on a very fine frequency grid (essentially realizing a continuum). In (b), both the
frequencies and the amplitude are sampled while keeping the total spectral weight unchanged.}
\label{Fig.sacpara1}
\end{figure} 

We here discuss SAC based on the most generic (unconstrained) form,
\begin{equation}\label{eq.somegadeltas}
S(\omega) = \sum_{i=1}^{N_\omega} a_i \delta(\omega-\omega_i),
\end{equation}
where the sampled spectral function consists of a collection of a large number
$N_\omega$ of $\delta$-functions (typically with $N_\omega$ of order $10^3$ or larger) with amplitudes $a_i$ at frequencies $\omega_i$ effectively in the
continuum; in practice a grid with a number of points orders of magnitude larger than $N_\omega$ with spacing $\Delta_\omega\sim10^{-5}$. The reason for
sampling $\delta$ functions on this grid instead of the continuum (with machine precision $\Delta_\omega$) is efficiency, to be able to use a precomputed
kernel while still effectively sampling the continuum. The spectral weight is averaged in the form of a histogram with bin width much larger than 
$\Delta_\omega$ but small enough to capture the details of the spectrum. Here we typically use a bin size of $\sim 10^{-3}$.

The spectrum is sampled with a Boltzmann-like probability distribution,
\begin{equation}\label{P(S)}
 P(S|\bar G) \propto \text{exp}\left (-\frac{\chi^2}{2\Theta}\right ),
\end{equation}
with $\chi^2/2$ playing the role of the energy of the system at a fictitious temperature $\Theta$, the value of which is determined in a way discussed
further below. We consider two versions of unconstrained SAC, one where the amplitudes are fixed uniformly at $a_i=1/N_\omega$ and another where the
amplitudes are also sampled along with the frequencies $\omega_i$. These two parametrizations of the spectral function are illustrated in
Fig.~\ref{Fig.sacpara1}. These parametrizations correspond to slightly different entropies \cite{Shao23,Ghanem23}.

We use the SSE method to generate statistically unbiased imaginary-time correlation function data $\bar G_i \equiv \bar G(\tau_i)$ for
a set of time points $\tau_i$, using $\beta \sim L$ or $2L$ for sufficiently converged ground state results; in the results reported here,
we have not detected any statistical differences between results at these two choices of $\beta$. We denote the conventional statistical
error (one standard deviation of the mean) by $\sigma_i$. However, since the fluctuations at different time points are correlated, the SAC
procedure requires the use of the full covariance matrix with elements $C_{ij}$ defined using bin averages $G^b_i$ for a large number of
bins $b = 1,\ldots N_B$:
\begin{equation}
 C_{ij}=\frac{1}{N_B(N_B-1)}\sum_{b=1}^{N_B} (G^b_i-\bar G_i)(G^b_j-\bar G_j).
\end{equation}
For a given spectral function $S(\omega)$ (an individual sample in the SAC process), the goodness of the fit to the QMC data is then given by
\begin{equation}\label{eq.chi2cov}
\chi^2=\sum_i \sum_j (G_i-\bar G_i)[C^{-1}]_{ij} (G_j-\bar G_j),
\end{equation}
where $G_i$ is obtained from $S(\omega)$ using Eq.~(\ref{eq.Gtau}). In practice, we transform the correlation functions to the eigenbasis of
the covariance matrix for a simpler form
\begin{equation}\label{eq.chi2}
\chi^2=\sum_i \frac{(G_i-\bar G_i)^2}{\epsilon_i},
\end{equation}
where $\epsilon_i$ is the $i$th eigenvalue of the covariance matrix and both $G_i$ and $\bar G_i$ now correspond to the respective
components after transformation to the new basis.

We use a $\tau$ grid with uniformly spaced points up to $\tau=2$ and then with the spacing increasing roughly linearly with $\tau$ for
$\tau > 2$. This choice reflects the fact that the relative statistical errors grow with $\tau$, and, therefore, the correlated data points
provide less independent information with increasing $\tau$. For a given wavenumber $k$, above some value of $\tau$ the data are completely
noise dominated, and we therefore impose a cutoff to exclude points for which the relative statistical error exceeds $10\%$. In
Ref.~\cite{Shao23}, an error level was defined as the statistical error (conventional standard deviation) of the normalized
$G(\tau)$ at the cutoff $\tau$. For all results reported here, the error level was better than $10^{-5}$.

When sampling according to the probability distribution in Eq.~\eqref{P(S)}, lowering $\Theta$ will result in a lower mean goodness of fit $\langle\chi^2 \rangle$. An optimal $\Theta$ value corresponds to balancing both goodness of fit (ensuring a statistically acceptable $\langle\chi^2 \rangle$ value) and entropy (avoiding over-fitting to statistical noise in $G$). A simple criterion has been developed to satisfy this balance, by requiring that $\langle\chi^2 \rangle$ is above the
minimum goodness of fit $\chi^2_\text{min}$ by roughly one standard deviation of the corresponding $\chi^2$ distribution;
\begin{equation}\label{eq.criterion}
\langle\chi^2(\Theta)\rangle=\chi^2_\text{min}+a\sqrt{2\chi^2_\text{min}},
\end{equation}
where $a$ is of order unity (typically we take $a \approx 0.5$). Here $\chi^2_\text{min}$ also serves as a proxy for the effective
number of degrees of freedom of the fit, $N_{\rm dof}$, based on the expectation that $\chi^2_{\rm min} = N_{\rm dof}$. We determine
$\chi^2_\text{min}$ to sufficient accuracy by a simulated annealing procedure in which $\Theta$ is gradually reduced until $\langle\chi^2(\Theta)\rangle$
converges, after which $\Theta$ is adapted to satisfy Eq.~(\ref{eq.criterion}) and the average spectrum is accumulated. We recently demonstrated
that this criterion also matches well the optimum value defined using statistical cross-validation \cite{Schumm24}. 

With the fixed-amplitude parametrization, the SAC method becomes equivalent to the standard maximum-entropy method (MEM) \cite{Jarrell96} in the
limit of large $N_\omega$, if the MEM entropy weighting factor is chosen such that the goodness of the fit of the solution equals $\langle \chi^2\rangle$
produced by the SAC method. If the amplitudes are also sampled, SAC becomes equivalent to a generalized MEM method with an entropic weighting different
from that of the conventional Shannon entropy \cite{Shao23,Ghanem23}. Thus, with typical QMC data there can be notable differences in the averaged $S(\omega)$
produced using these two unconstrained parametrizations. We recently demonstrated that cross-validation can be used to discriminate between different
parametrizations \cite{Schumm24}.

The main advantage of the SAC scheme is that various constraints can be introduced that produce sharp spectral features that are not always easy
to incorporate within the MEM scheme. We will apply and optimize constraints in the later sections, and finally use cross-validation to determine the optimal constrained parametrization.

\section{Ordered phases}
\label{sec:SpectrumAFM}

Here we study the dynamic structure factor in the ordered AFM and SAFM phases. We employ different SAC parametrizations, first with the
basic unconstrained parametrizations in Fig.~\ref{Fig.sacpara1} and later with various constraints to improve the resolution of the
sharp magnon peak. We here will study chains of length $L=256$ to test the different parametrizations and elucidate the basic aspects of
$S(k,\omega)$; finite-size effects will be mostly addressed in later sections.

\subsection{Unconstrained SAC in the AFM phase}

We begin in the AFM phase, where previous SWT calculations have demonstrated a nonlinear dispersion relation in the limits $k \to 0$ and
$k \to \pi$ \cite{Yusuf04,Laflorencie05}. The previous tDMRG calculations~\cite{Yang21} showed some spectral weight also above the magnon peak, which
is absent in linear SWT.

\begin{figure}
\includegraphics[width=8.3cm]{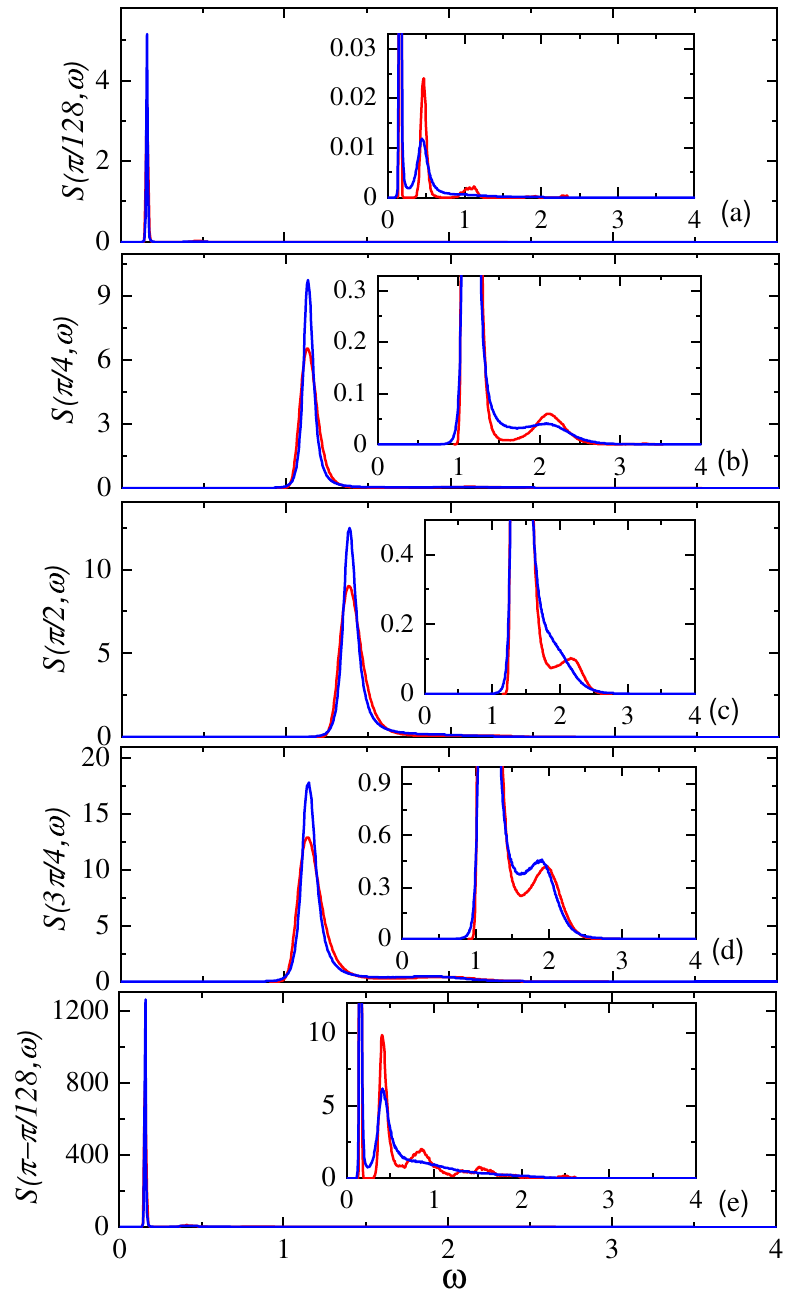}
\caption{Dynamic structure factor for an $L=256$ chain at five different momenta in the AFM phase at $\alpha=2$. The blue and red curves were obtained
by sampling with and without amplitude updates, respectively. In the  main panels the vertical scale has been adapted to focusing on the dominant magnon peak,
while the insets show details of the continuum above the magnon peak.}
\label{unconstafm}
\end{figure}

We show examples of $S(k,\omega)$ obtained using SAC with both unconstrained parametrizations in Fig.~\ref{unconstafm}. Here the long-range
exponent is $\alpha=2$, which is sufficiently below the QLRO transition point $\alpha_c \approx 2.23$ for the system to exhibit robust AFM order.
The dominant magnon peak is somewhat sharper when amplitude updates are included, but the location of the peak is essentially consistent among the
two parametrizations. Continua above the main peaks are also apparent. For the $k$ points closest to $0$ and $\pi$, the continua (which have a very small
relative weight) show a peak structure that is more pronounced with the constant-amplitude parametrization. Such ``ringing'' behavior is a common
artifact of numerical analytic continuation and can often be traced to the inability of unconstrained methods (including also MEM) to reproduce sharp low-frequency features; the distortions at the lower edge then lead to compensating
distortions at higher frequencies to match the imaginary-time correlations within statistical errors.

Another notable feature of Fig.~\ref{unconstafm} is that the main peak is narrower at lower energy. We will discuss the shape of the magnon peak
in detail in later sections, but here already note that the structure factor should evolve toward a pure $\delta$-function, $S(k,\omega) = a_k \delta(k-\omega_k)$,
for $k \to 0,\pi$. The narrowing of the dominant magnon peak is therefore expected, but the peak width cannot be expected to be fully captured by SAC with
unrestricted sampling, especially at higher energies where there is a larger continuum above the main peak. It is well known, and clear from Eq.~(\ref{eq.Gtau}),
that low-energy features effectively produce more information in the imaginary-time correlation function $G(k,\tau)$, given that the decay with $\tau$
is slower and, therefore, longer times $\tau$ can be reached before the data become noise dominated. Considering these limitations, results such as those in Fig.~\ref{unconstafm} can still be used to
extract the magnon dispersion relation to reasonable precision and give some quantitative information on a continuum above the magnon peak. Overall, the spectral curves are slightly better when amplitude updates are included since output profile has sharper magnon peak and less artificial ``ringing'' effects. However, the detailed shapes of the spectra are not reliable using these two parametrizations, and much better results can be reproduced using the constraints introduced below.

\begin{figure}
\includegraphics[width=8.3cm]{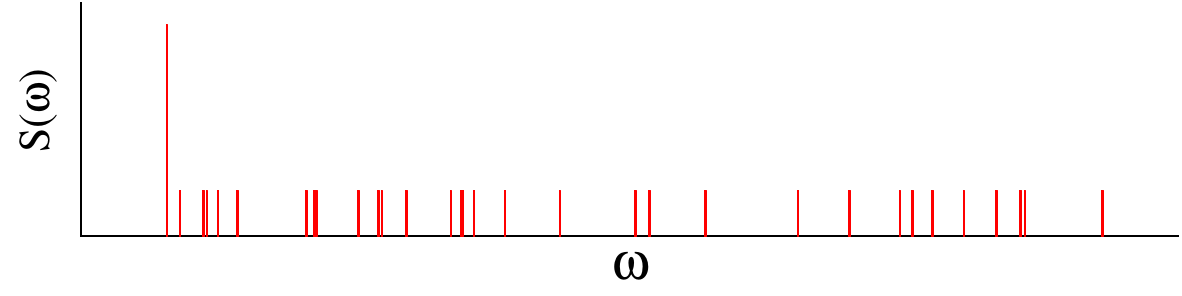}
\caption{Parametrization of a spectral function $S(\omega)$ with an edge consisting of a macroscopic $\delta$-function whose location $\omega_0$ serves as the
lower bound for a large number equal-amplitude microscopic $\delta$-functions representing a continuum. The location $\omega_0$ is sampled along with the other locations
$\omega_{i>0} > \omega_0$. The optimal weigh $a_0$ of the macroscopic $\delta$-function is determined by a scan such as that in Fig.~\ref{a0scan}.}
\label{singlepeakpara}
\end{figure} 

\subsection{Sharp magnon peak in the AFM phase}
\label{sec:afmdelta}

The long-range interacting Heisenberg chain in the AFM phase likely shares properties with the standard 2D Heisenberg model, even though the magnon
dispersion is anomalous in the former. High-order SWT calculations for the 2D Heisenberg model
\cite{Igarashi92,Canali93,Igarashi05,Syromyatnikov10,Powalski15,Powalski18}
show that the dynamic structure factor contains a dominant $\delta$-function at the lowest frequency $\omega_k$ (the dispersion relation), followed by
an incoherent multi-magnon continuum,
\begin{equation}\label{eq:Sqw1}
 S(\textbf{k},\omega)=S_0(\textbf{k})\delta(\omega-\omega_\textbf{k})+S_c(\textbf{k},\omega)\theta(\omega-\omega_k),
\end{equation}
where $S_0(\textbf{k})$ is the spectral weight of the leading magnon pole and $S_c$ represents the continuum weight above $\omega_k$ (as indicated
by the step function $\theta$). Some broadening of the magnon peak may also be expected. Within SWT, for collinear antiferromagnets broadening will appear
at very high order in $1/S$ (while appearing at low order in non-collinear systems) \cite{Chernyshev06}, but there are no specific quantitative calculations
including this aspect, as far as we are aware.

A SAC parametrization motivated by Eq.~(\ref{eq:Sqw1}) was implemented in a study of the two-dimensional (2D) Heisenberg model \cite{Shao17} and we here use the same
approach for the long-range chain in the AFM phase. In this SAC ``$\delta$-edge'' parametrization, illustrated in Fig.~\ref{singlepeakpara}, the normalized spectral function includes a ``macroscopic''
$\delta$-function of weight $a_0$,
\begin{equation}
a_0(k)=\frac{S_0(k)}{\int d\omega S(k,\omega)},
\label{a0def}
\end{equation}
and a continuum above its energy $\omega_0$. The continuum is parametrized in the same way as in the equal-amplitude case,
but now with the constraint that the $N_\omega$ ``microscopic'' $\delta$-functions are only allowed above $\omega_0$. Their total spectral weight is accordingly $(1-a_0)$, which is divided evenly, i.e. their microscopic weights are $(1-a_0)/N_\omega$.

\begin{figure}
\includegraphics[width=8cm]{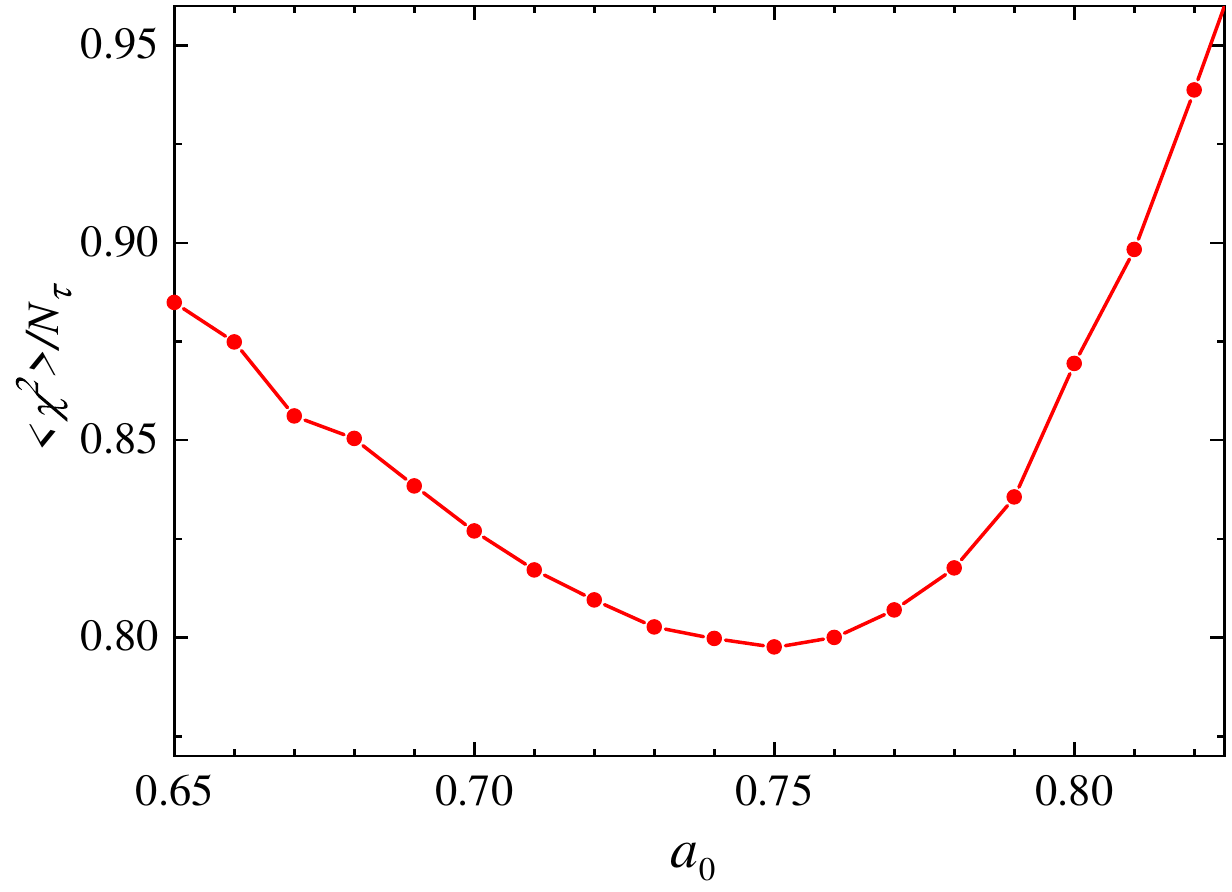}
\caption{Scan over the macroscopic $\delta$-function weight $a_0$ with the SAC parametrization illustrated in Fig.~\ref{singlepeakpara}, here for the
$L=256$ chain at $\alpha=2$ and $k=\pi/2$. The $\langle \chi^2\rangle$ minimum represents the optimal weight. Here the sampling temperature $\Theta$ was chosen
such that the criterion Eq.~(\ref{eq.criterion}) is satisfied with $a\approx 0.5$.}
\label{a0scan}
\end{figure} 

As argued in Refs.~\onlinecite{Shao17} and \onlinecite{Shao23}, at a fixed sampling temperature $\Theta$, increasing $a_0$ from $0$ leads to a reduction in
configurational entropy (with its associated tendency to spread out the spectrum below the actual edge), thus leading to a better fit and
decreasing $\langle \chi^2\rangle$. However, when $a_0$ exceeds its correct value (under the assumption that the edge really is a $\delta$-function
or a very narrow peak) the spectrum can no longer fit the imaginary-time data; hence, $\langle \chi^2\rangle$ will increase sharply above some value
of $\alpha_0$. Thus, a minimum in $\langle \chi^2\rangle$ is observed when compared across a series of constant-$a_0$ SAC runs. Detailed tests
with synthetic data in Ref.~\onlinecite{Shao23} showed that the location of the minimum approached the true $a_0$ value as the data quality is
improved, with the the sampling temperature adapted so that the minimum $\langle \chi^2\rangle$ roughly satisfies the criterion Eq.~(\ref{eq.criterion}).
Figure~\ref{a0scan} shows an example of such an $a_0$ scan, using the same SSE data as previously in Fig.~\ref{unconstafm}(c).

\begin{figure}
\includegraphics[width=8.4cm]{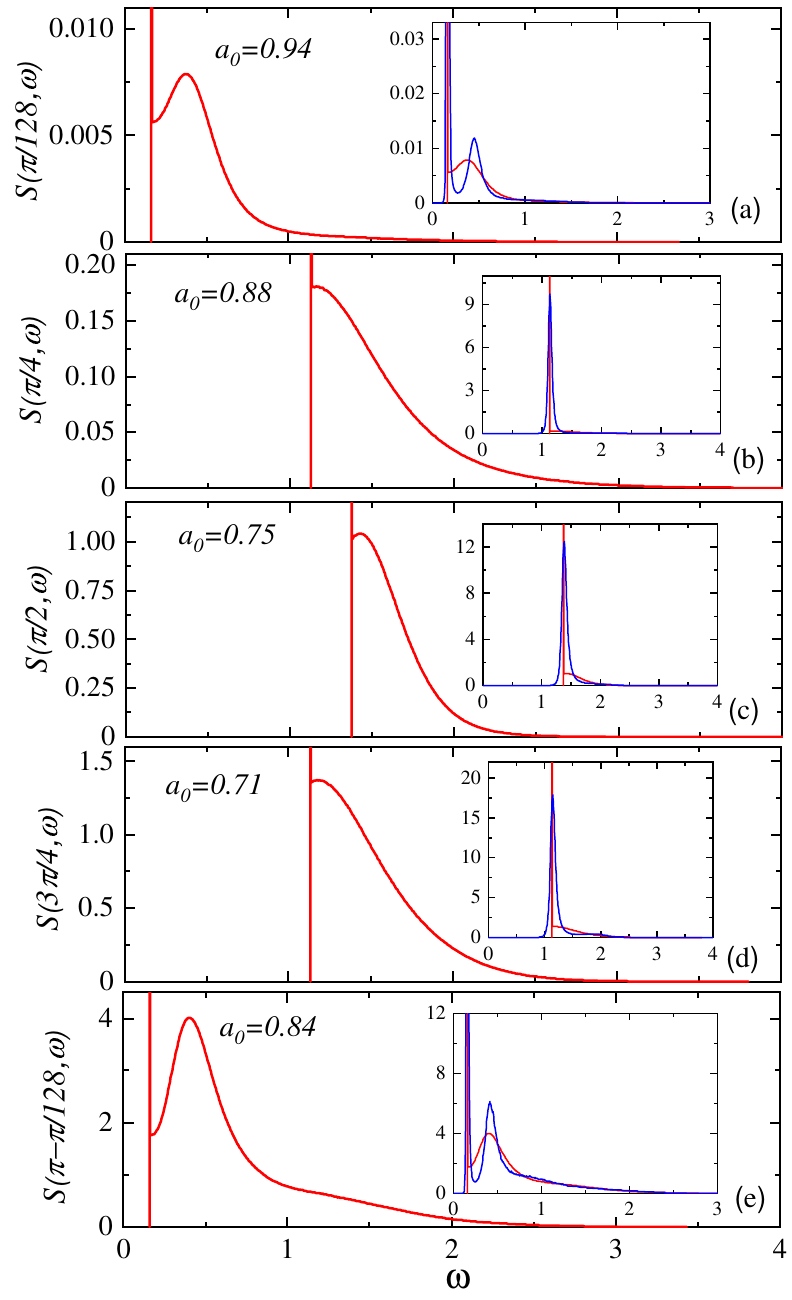}
\caption{Dynamic structure factors at five different momenta for the $L=256$ chain with $\alpha=2$ obtained with the $\delta$-edge parametrization in
Fig.~\ref{singlepeakpara} and with the weight $a_0$ optimized as in Fig.~\ref{a0scan}. The optimal $a_0$ value is indicated in each panel. The $k$ points are
the same as in Fig.~\ref{unconstafm} and the results obtained there with amplitude updates are compared with those with the $\delta$-edge in the insets.}
\label{Fig.a2repsS(k,w)}
\end{figure} 

Figure \ref{Fig.a2repsS(k,w)} shows results for the same system and $k$ points as in Fig.~\ref{unconstafm}. Here the weight of the leading $\delta$-function is always
relatively large and its location coincides closely with that obtained by unrestricted sampling, as shown in the insets of Fig.~\ref{Fig.a2repsS(k,w)}. The spectral
weight in the continuum above the magnon peak falls roughly within the same energy range as before, but the details of the different profiles are clearly different.

\begin{figure}
\includegraphics[width=8.4cm]{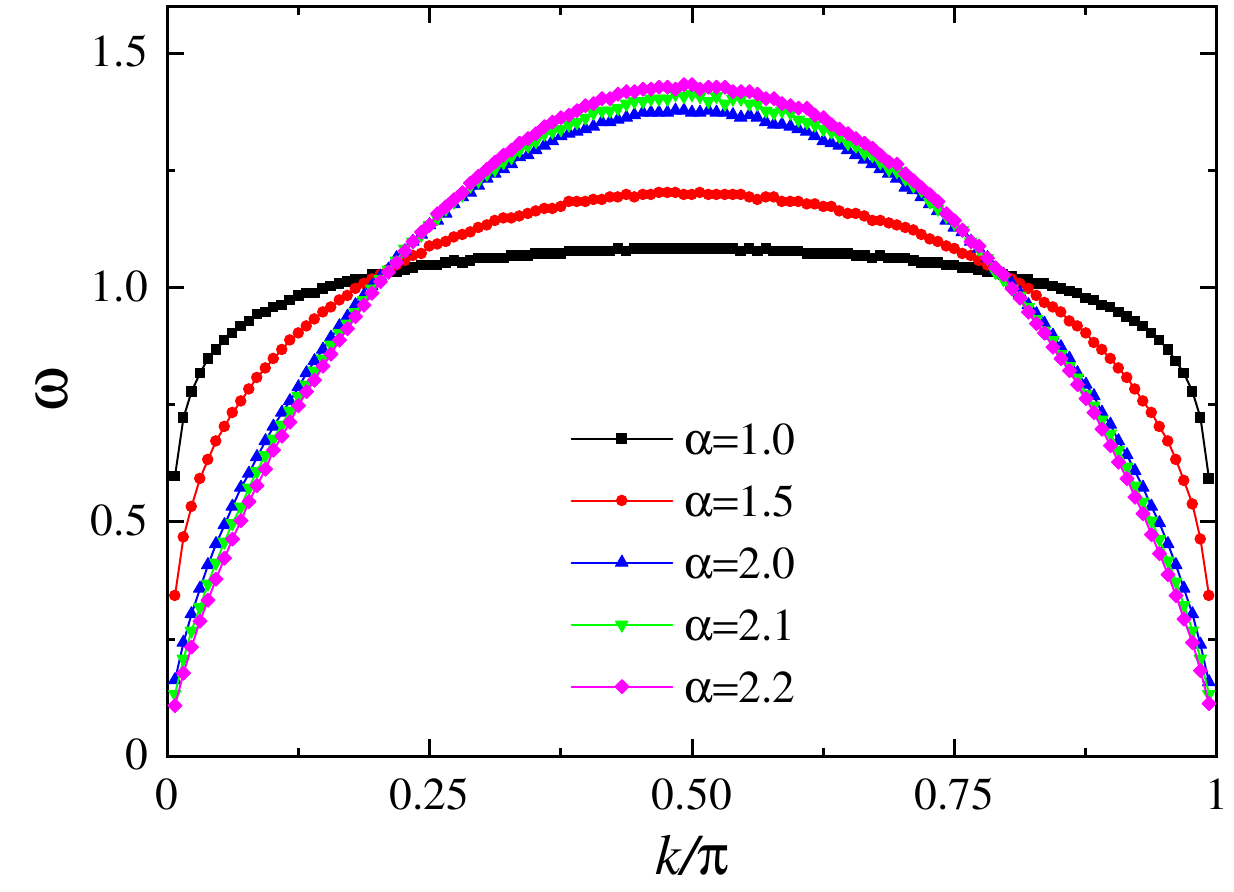}
\caption{Dispersion relations $\omega_k$ for different long-range interaction exponents $\alpha$ in the AFM phase for $L=256$. Each point is extracted
from the position of the $\delta$ function in spectral functions exemplified in Fig.~\ref{Fig.a2repsS(k,w)}.}
\label{Fig.w_vs_k}
\end{figure}

It should be noted that the magnon location $\omega_0$ does fluctuate in the SAC process, but not enough to provide visible broadening on the energy scale used in
Fig.~\ref{Fig.a2repsS(k,w)}. In the limit of large $N_\omega$, the fluctuations about the optimal location vanish completely \cite{Shao23} and the calculations shown
here are in practice close to that limit. The dispersion relation obtained from the magnon peak is graphed in Fig.~\ref{Fig.w_vs_k} for several values of $\alpha$ between
$1$ and $2.2$. Here the nonlinearity close to $k=0$ and $\pi$ is obvious for the smaller $\alpha$ values. There are still some finite-size effects in these results,
especially for the smaller $\alpha$ values, which we will analyze in more detail in Sec.~\ref{sec:dispersionAFM}, where we extract the exponent $z$ of the dispersion
relation $\omega_k \propto k^z$ and compare it with SWT results.

\begin{figure}
\includegraphics[width=8.4cm]{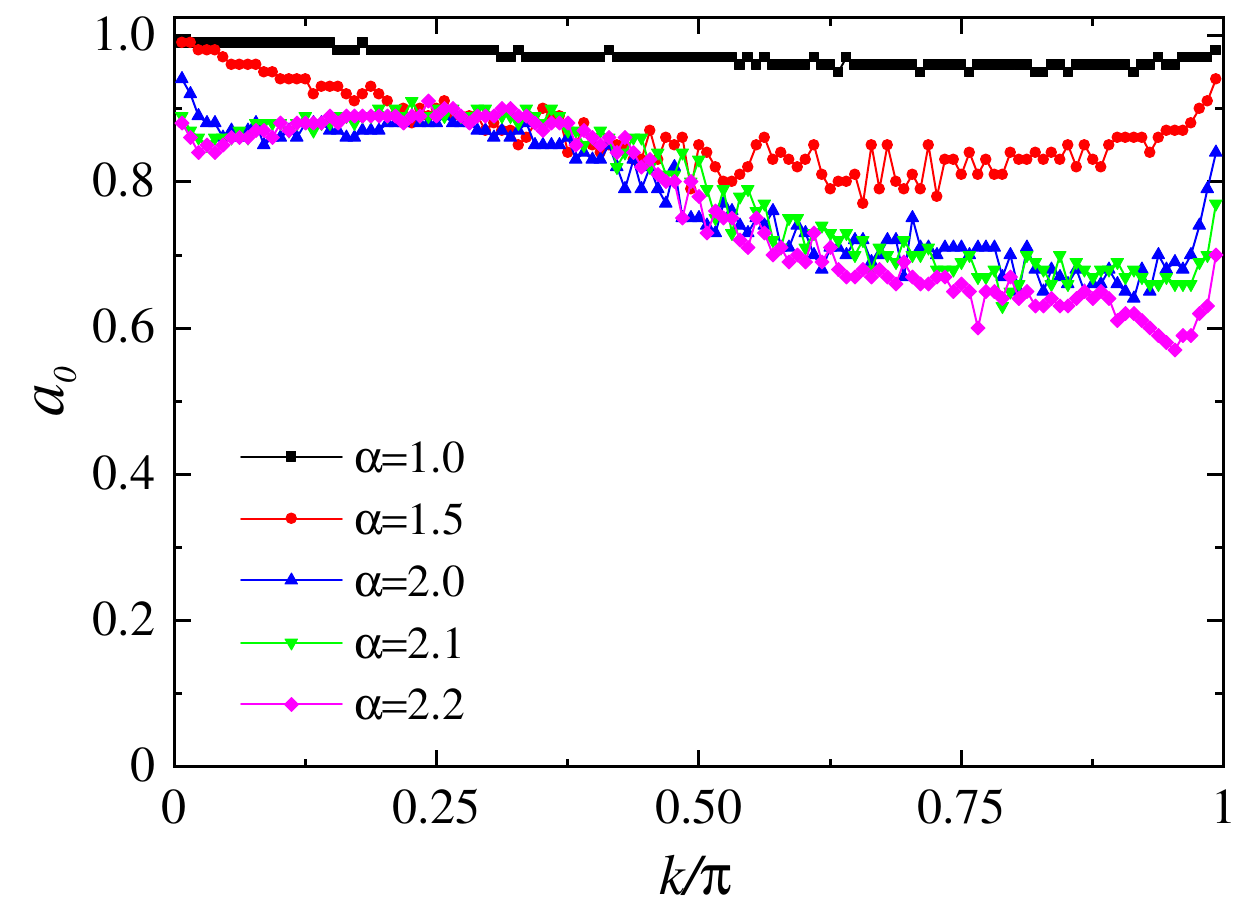}
\caption{Relative weight $a_0$ of the single-$\delta$ magnon peak vs the momentum $k$ for the same cases as the dispersion relations presented
in Fig.~\ref{Fig.w_vs_k}.}
\label{Fig.a0_vs_k}
\end{figure}

The relative magnon weight $a_0$ is graphed versus $k$ for representative $\alpha$ values in Fig.~\ref{Fig.a0_vs_k}. In some cases there are relatively large
variations in $a_0$ for nearby values of $k$, which reflects the statistical uncertainty in the optimal $a_0$ value. Similarly to the 2D case \cite{Sandvik01,Shao17},
the single-magnon contributions should exhaust the spectral weight for $k \to 0$ and $\pi$. For $\alpha > 1$, there is indeed a sharp increase in $a_0$ as these
limits are approached, but larger systems would be required to study in detail how $a_0 \to 1$. Note that, while the dispersion relations are very symmetric about
$k=\pi/2$ in Fig.~\ref{Fig.w_vs_k}, the magnon weight has no such symmetry in Fig.~\ref{Fig.a0_vs_k}, with $a_0$ being much larger for $k < \pi/2$. This difference in
the neighborhood of the two gapless points is also a known feature in the 2D Heisenberg model.

For the larger $\alpha$ values, close to the AFM--QLRO transition at
$\alpha \approx 2.23$, there is a more complex shape of $a_0(k)$, with a local maximum in the weight at $k/\pi \approx 0.3$, where $a_0$ even exceeds the value at
some smaller $\alpha$, exemplified in Fig.~\ref{Fig.a0_vs_k} by $\alpha=1.5$. Upon decreasing $\alpha$, the magnon weight approaches unity for all $k$ in
Fig.~\ref{Fig.a0_vs_k}, which is not surprising, given that the spectrum should consist only of a $\delta$-peak in the SAFM phase (as we will show explicitly
below in Sec~\ref{sec:SpectrumSAFM}) entered at still smaller $\alpha$.

We should point out here that the magnon weight $a_0$ obtained with the $\delta$-edge parametrization can be expected to be correct only if the
continuum is well modeled by the mean density of the microscopic $\delta$-functions illustrated in Fig.~\ref{singlepeakpara}. In practice, this
implies that the continuum itself must not be too strongly peaked (is not close to divergent) at the edge---as seen in Fig.~\ref{Fig.a2repsS(k,w)},
the continuum produced by this SAC parametrization is indeed not strongly peaked at the edge. If the true spectrum has a very sharp peaked continuum,
some of its weight close to the edge will be effectively included in $a_0$. As we will discuss in later sections, the smoothness condition on the continuum
is likely violated close to the QLRO boundary at $\alpha_c=2.23$, and an improved parametrization of the continuum then should be used to obtain a better
estimate of the true value of $a_0$ for given $k$. The results presented in Fig.~\ref{Fig.a0_vs_k} nevertheless represent a useful measure of the
accumulation of spectral weight very close to the edge as $k$ is varied.

\subsection{Exact magnon eigenstates in the SAFM phase}
\label{sec:SpectrumSAFM}

\begin{figure}
\includegraphics[width=8.3cm]{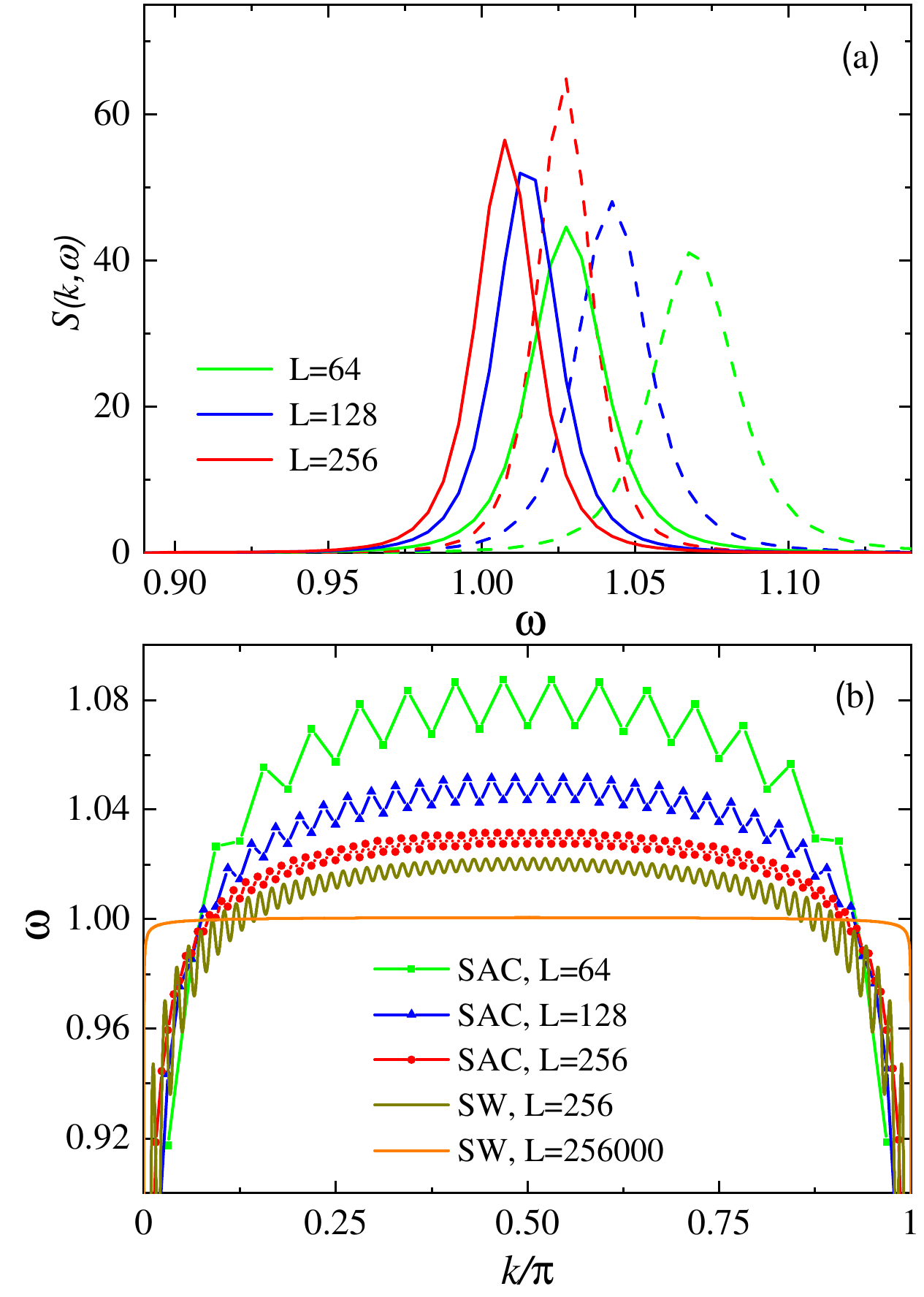}
\caption{SAC and SWT results in the SAFM phase at $\alpha=0.5$. In (a) the dynamic structure factor obtained with unconstrained SAC for system sizes $L=64$, $128$,
and $256$ are shown for $k=\pi/8$ (solid curves) and $\pi/2$ (dashed curves). In (b), the dispersions relation obtained for the same system sizes with the
$\delta$-edge parametrization are compared with SWT results for $L=256$ and $256000$; details of the SWT calculations are presented in Appendix \ref{appsec:swa1a2}.}
\label{Fig.safm}
\end{figure}

The SAFM phase at $\alpha \le 1$ is characterized by the order parameter taking its maximum possible value, $m_s=1/2$, in the thermodynamic limit.
In linear SWT, the magnons are dispersionless in the SAFM phase, i.e., the dynamic exponent $z=0$, which corresponds to localized
excitations. We expect that the spectral function here consists solely of a $\delta$-function, with no broadening or continuum above it. With
unrestricted SAC, we indeed obtain very sharp peaks, as exemplified by the $\alpha=0.5$ results at momenta $k=\pi/8$ and $k=\pi/2$
in Fig.~\ref{Fig.safm}(a) for system sizes $L=64$, $128$, and $256$ [note that the scale of the $x$-axis in Fig.~\ref{Fig.safm}(a) is enlarged
compared to Figs.~\ref{unconstafm} and~\ref{Fig.a2repsS(k,w)}]. The peaks narrow somewhat with increasing $L$ and are nearly symmetric, which is
suggestive of no significant continuum above the magnon. The peak position shifts slightly toward lower energy with increasing $L$. Most likely, these are
resolution-limited peaks, with the narrowing due to better resolution at lower energy. Indeed, we can also fit the data to a single $\delta$ function
($a_0=1$) whose frequency is optimized, with excellent goodness of fit. On statistical grounds no other features should then
be included in modeling the spectral function.

Figure~\ref{Fig.safm}(b) shows the dispersion relation obtained in fits solely to a $\delta$-function. The finite-size effects are very large here, and we also
show SWT results (computed according to Refs.~\onlinecite{Yusuf04,Laflorencie05} with a small adaptation to the precise form of our Hamiltonian, as detailed in
Appendix \ref{appsec:swa1a2}). In our SWT calculations we observe that $L=256$ is not a sufficiently large size to remove finite-size-effects, and even with a system
$10^3$ times as large, there are still visible deviations from the
constant form, most notably close to $k=0$ and $\pi$. The SAC results show the same oscillating behavior versus $k$ as the SWT curves for small $L$. We will not attempt
to reach convergence in the system size with the SAC in this case, as clearly prohibitively large systems would be required. As will be shown in
Sec.~\ref{sec:dispersionAFM} and Appendix \ref{appsec:swa1a2}, convergence is much faster in the AFM phase, and for $\alpha \ge 1.5$ we are able to extract reliable
results for the dynamic exponent $z$ governing the $k \to 0,\pi$ limits of the dispersion relation.

\subsection{Broadened magnon peak in the AFM phase}
\label{subsec:sacNdelta}

\begin{figure}[t]
\includegraphics[width=8.3cm]{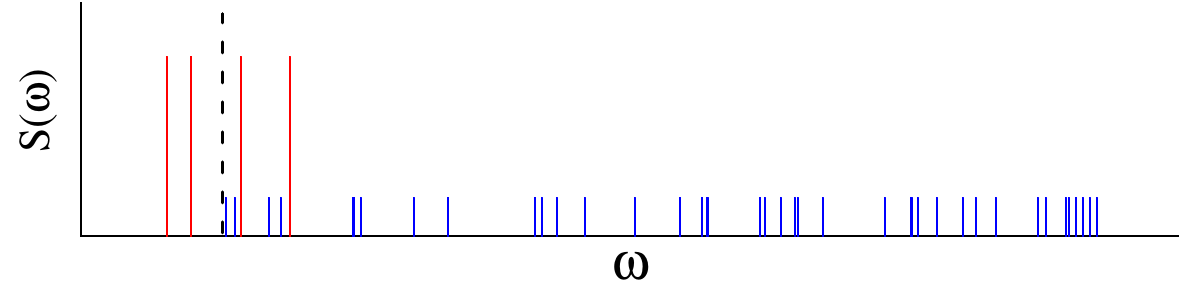}
\caption{Parametrization of the spectral function with a relatively small number $N_p$ of $\delta$-functions representing a broadened quasi-particle peak (red bars)
  of total relative weight $A_0$, and with a continuum of weight $1-A_0$ represented
  by a number $N_c \gg N_p$ of equal-weight $\delta$-functions (blue bars). The continuum
contributions are normally lower-bounded by the first moment of the quasi-particle contributions.}
\label{Fig.sacpara3}
\end{figure}

In Eq.~(\ref{eq:Sqw1}) the magnon peak was assumed to be a single $\delta$ function, but in reality it is possible that this feature is somewhat broadened,
as has been discussed in the case of the 2D Heisenberg model \cite{Powalski18,Chernyshev06}, though even there it is difficult to reliably compute the width of
the peak (and we are not aware of any specific results). Quasi-particle broadening should be considered as distinct from the continuum extending beyond the
peak, though both forms of
deviations from a sharp ($\delta$-function) peak will be the result of interactions. In order to attempt to extract the quasi-particle width of the long-range
model, we next employ SAC with a multi-$\delta$-peak parametrization \cite{Shao23}, illustrated in Fig.~\ref{Fig.sacpara3}. We here focus on our main results
with this parametrization and present details of the procedures (with minor extensions of Ref.~\cite{Shao23}) in Appendix~\ref{app:multipeak}. We note from the
outset that our calculations described below do not indicate a standard peak broadening, i.e., the parametrization used here is not appropriate for describing
the true spectral function. The steps leading to this conclusion are nevertheless useful as an example of how to test different parametrizations when there
is not sufficient prior knowledge of the spectral function and the goal is to narrow the range of possibilities.

\begin{figure}
\includegraphics[width=8.3cm]{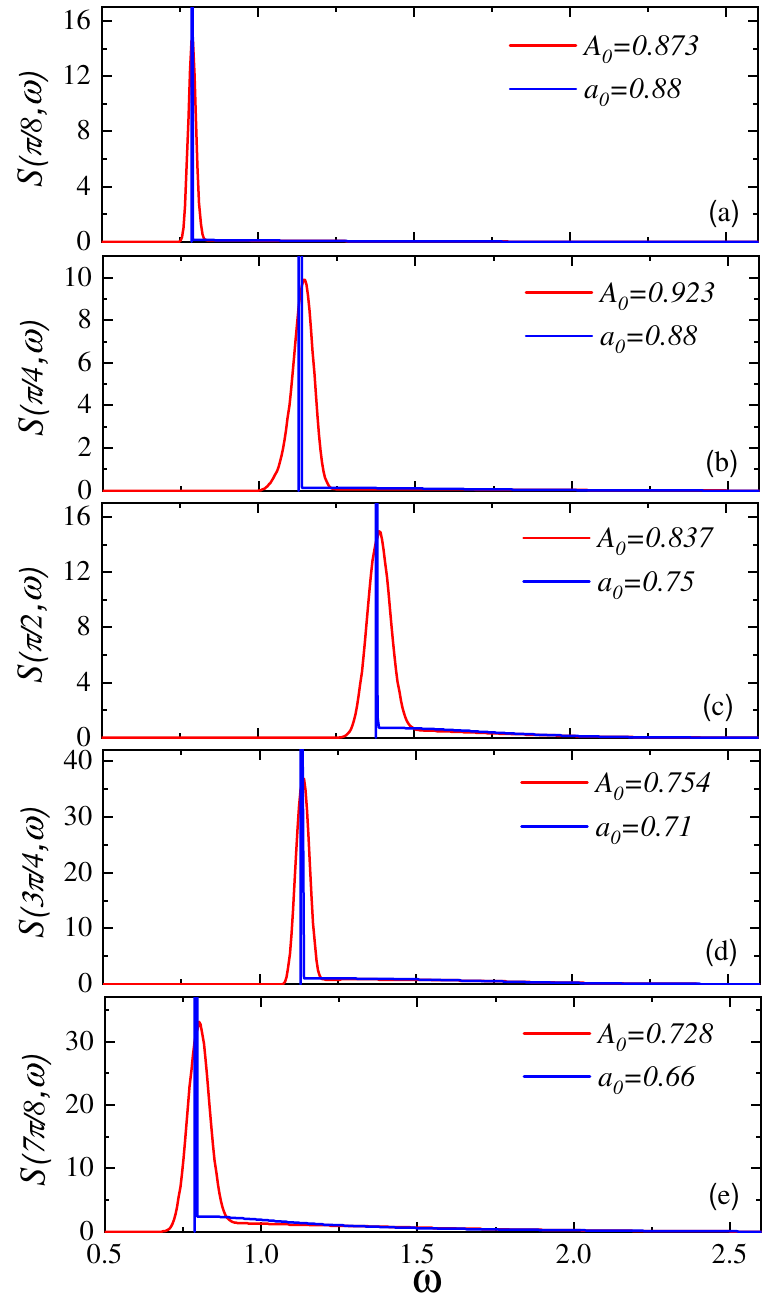}
\caption{Dynamic structure factor of the $L=256$ chain with $\alpha=2$ at five different momenta obtained by SAC with the multi-$\delta$ peak parametrization
illustrated in Fig.~\ref{Fig.sacpara3}. The results are graphed along with those of the single-$\delta$ edge parametrization. The corresponding relative
quasi-particle weights $A_0$ (multi-$\delta$) and $a_0$ (single-$\delta$) are indicated in each panel.}
\label{Fig.a2repsS(k,w)width}
\end{figure} 

\begin{figure}
\includegraphics[width=8.3cm]{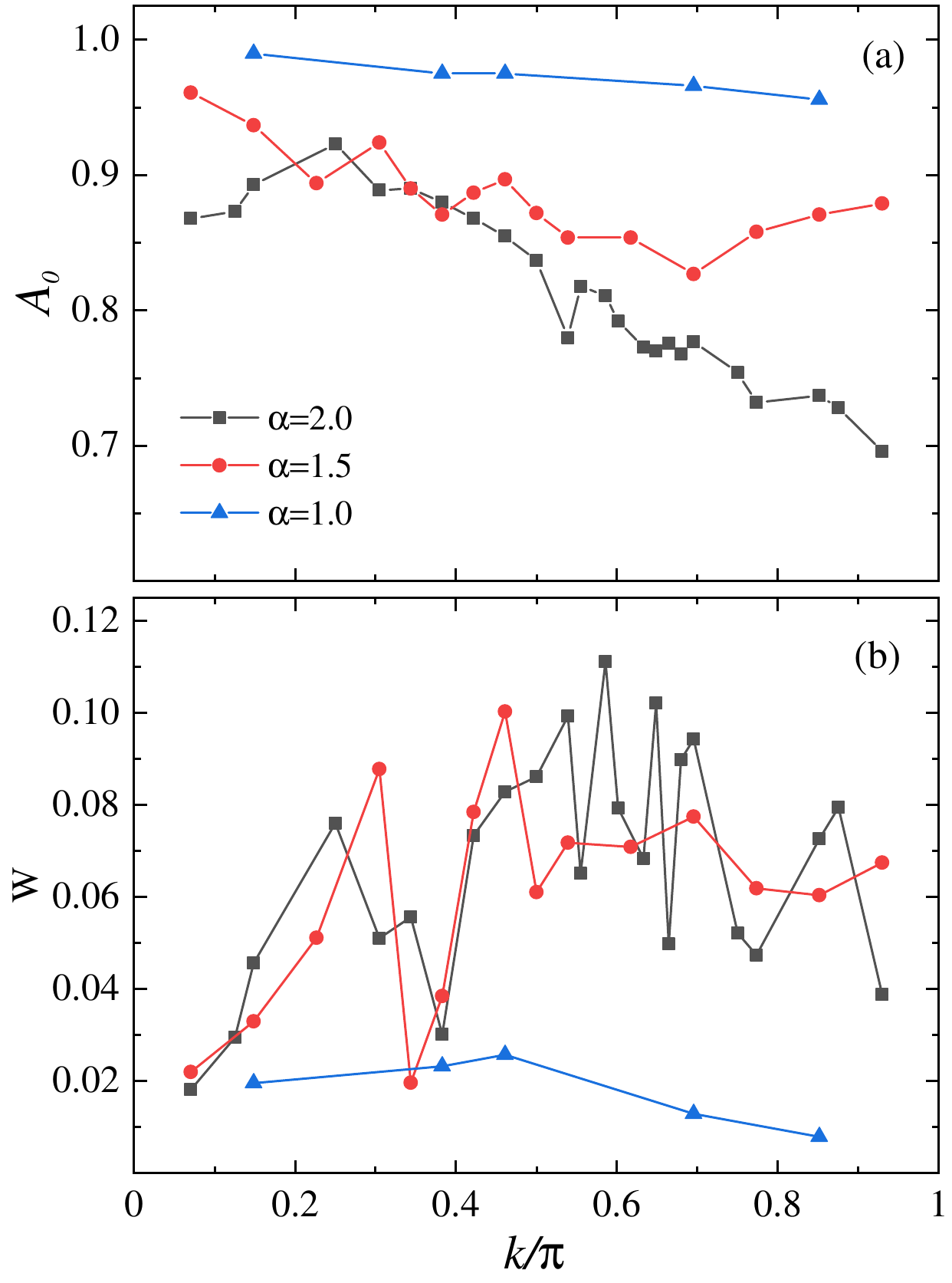}
\caption{Amplitude (a) and full-width at half-height (b) of the broadened magnon peak versus momentum $k$ for $\alpha=1.0$, $1.5$ and $2.0$, obtained
in SAC calculations as those exemplified by Fig.~\ref{Fig.a2repsS(k,w)width}.}
\label{width_combine}
\end{figure}

Applying the multi-$\delta$ peak SAC parametrization for some representative momenta $k$ for the $L=256$ chain at $\alpha=2$, we obtain the spectra shown in
Fig.~\ref{Fig.a2repsS(k,w)width}, with each panel indicating the corresponding total peak weight that we now denote as $A_0$ to distinguish it from the single-$\delta$
edge weight $a_0$. Spectra obtained using the single-$\delta$ peak parametrization are also included for comparison. The peak positions obtained with these two SAC
parametrizations match each other very well, even though the low-energy tails below the sharp single-$\delta$ peak imply a small compensating shift of the broadened
peak to higher frequency. The magnon amplitude is graphed versus $k$ Fig.~\ref{width_combine}(a) and shows similar trends as the single-$\delta$ amplitude
in Fig.~\ref{Fig.a0_vs_k}.

We define the width of the magnon peak as the total width at half of the maximum amplitude, which we plot for different $\alpha$ values and various $k$ points
in Fig.~\ref{width_combine}(b).
While these results are rather noisy, as is obvious from the large scatter among the data points, the trend of two maxima and
a rather sharp minimum between them is still clear for $\alpha=2$ and $\alpha=1.5$, especially given that the behaviors look quite similar for these two $\alpha$
values. The width is considerably smaller at $\alpha=1$, which is expected given that the system here is in or close to the SAFM phase with its pure $\delta$-function
spectrum. For $k$ close to $0$, there is a general trend that the magnon peak becomes sharper, matching our expectation that the spectrum in this limit
evolves into a pure $\delta$ function for all values of $\alpha$. This narrowing of the peak is expected also for $k\to \pi$, but the noisier data here makes
it more difficult to confirm explicitly.

A key question now is whether the multi-$\delta$ peak representation really delivers the correct peak shapes, or whether the spectra are significantly affected by
entropic broadening. The rather sharp minimum between the two maxima in Fig.~\ref{width_combine}(b) for $\alpha=2$ and $\alpha=1.5$ suggests a resolution of $0.02$
or better of the width, as this is the minimum width extracted using our method. Given that the peak location and $A_0$ evolve smoothly versus $k$, there is no
obvious reason why the peak width should exhibit such a sharp minimum unless this $k$ dependence is a real feature.

It is possible to test the resolution within the method itself \cite{Shao23}, and we present such tests in Appendix~\ref{app:multipeak}. The tests indicate that the
$\alpha=1.5$ and $2$ peak widths in Fig.~\ref{width_combine}(b) are resolution limited at the momentum of the local width minimum ($k/\pi \approx 0.35$), while for
$k$ slightly away from this region the width is correctly estimated (though the peak shape is likely not correct, as we will discuss further below). For $\alpha=1$,
the calculated width in Fig.~\ref{width_combine}(b) may always be resolution limited.

Although these results appear to support a significantly broadened quasi-particle peak, the actual shape of the peak is not necessarily correctly captured by the
multi-$\delta$ approach. A natural alternative to the Gaussian-like profile that we have obtained here is a sharp edge at a lower bound of the spectral weight.
Such an edge, with an associated power-law divergence as the edge is approached from above, is well known in the standard Heisenberg chain with only $J_1$
interactions, and can be expected throughout the QLRO phase. In that case, the shape is directly related to the fractionalization of the excitation into two
spinons, which is not expected in the AFM phase. A sharp edge, but without a $\delta$-function, cannot be excluded as a possible spectral form for anomalous
magnons in the AFM phase. Such an edge of the form $(\omega-\omega_k)^{-p}$ is not necessarily divergent but can be characterized by a negative exponent $p$.
In the next section we will explore parametrizations suitable for spectra with sharp power-law edges edges in both the QLRO and AFM phases.

\section{Spectral edge with power-law divergence}
\label{sec:sacpowerlaw}

Bethe-ansatz (BA) results \cite{Muller97,Caux05,Caux05_2,Pereira06,Caux06} for the pure antiferromagnetic Heisenberg chain show that the
dynamic structure factor should be an edge at energy $\omega_k$ with a divergent spectral density for $\omega \to \omega_k^+$. Within the approximation of only
two-spinon contributions, the divergence has the form \cite{Muller97}
\begin{equation}\label{edgeba}
S(k,\omega \to \omega_k^+) \propto \frac{\ln(\omega -\omega_k)}{(\omega -\omega_k)^{1/2}},
\end{equation}  
and it is believed that, this form also applies when the higher-order spinon contributions are included. We expect that throughout the QLRO phased in our model, $S(k,\omega)$ should also exhibit an edge with a similar form, while there will likely be differences at higher energy. Given the broadened quasi-particle peak found in
the previous section, it is also possible, in principle, that there is a sharp edge at the lowest energy in the AFM phase instead of a $\delta$-function or
broadened Gaussian-like shape, though not necessarily with a divergence at the edge.

The simplest way to guarantee a sharp edge within SAC is to simply impose a lower bound at some energy $\omega_0$. If the bound is placed where the actual
spectrum has significant weight, a sharp edge will be produced. If $\omega_0$ is too high, a good fit to the data cannot be obtained, while a very low bound
(below the region where unconstrained sampling produces no spectral weight) will have no impact on the sampled spectrum. Once the bound eliminates some spurious
low-energy weight, a reduction of sampling entropy will be reflected in a lowered $\langle \chi^2\rangle$ value versus $\omega_0$ when sampling at fixed $\Theta$, until
the fit rapidly deteriorates when the bound exceeds the actual spectral edge. Thus, a minimum forms in $\langle \chi^2\rangle$, as in Fig.~\ref{a0scan}, signaling
an optimal lower bound within the parametrization. The resulting spectrum will not contain a divergence for $\omega \to \omega_0^+$, however. Note also that the edge
location $\omega_0$ cannot be sampled, because entropy will then push the edge to low energy, producing results identical to unconstrained sampling.

\begin{figure}
\includegraphics[width=8.3cm]{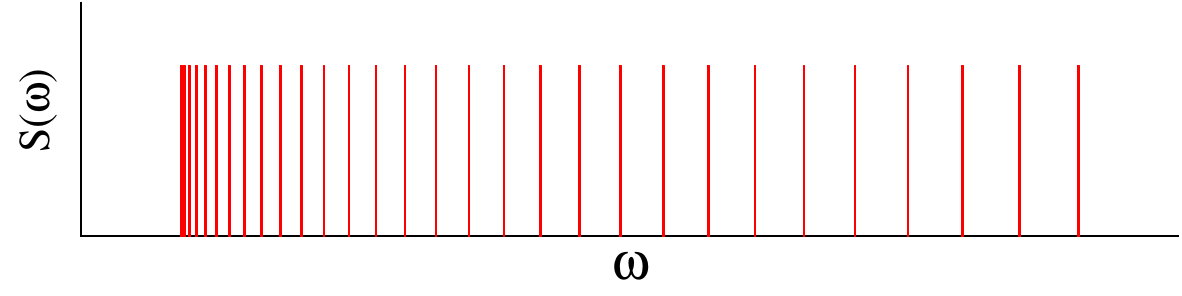}
\caption{SAC parametrization with equal-weight $\delta$-functions under the constraint of monotonically increasing spacing;
$\omega_i-\omega_{i-1}\leq \omega_{i+1}-\omega_i$. The position $\omega_0\equiv \langle \omega_{1}\rangle$ of the edge is not fixed (though it can
be if $\omega_0$ is known) but is also sampled with the constraint maintained.}
\label{Fig.sacpara4}
\end{figure} 

In order to produce a spectrum with power-law divergent edge in SAC, the constraint of a monotonically increasing distance between adjacent $\delta$ functions
can be imposed, i.e. $\omega_i-\omega_{i-1}\leq \omega_{i+1}-\omega_i$ for $i=1,\ldots,N_\omega$ \cite{Shao23}. The spectrum can then be defined using a self-generated grid based on the mean locations $\langle \omega_i\rangle$
and amplitudes $a_i$, as a better alternative to collecting spectral weight in a predefined histogram;
\begin{equation}
S(\omega_{i+1/2})=\frac{1}{2}\frac{a_{i}+a_{i+1}}{\langle \omega_{i+1}-\omega_{i}\rangle},
\end{equation}
where $\omega_{i+1/2}=(\langle \omega_{i}+\omega_{i+1}\rangle)/2$. In the simplest case, illustrated in Fig.~\ref{Fig.sacpara4}, the $\delta$-functions again
have equal amplitudes $a_i$ and the native entropic pressure in the presence of the monotonicity constraint then favors a singularity of the form
(on the self-generated grid)
\begin{equation}\label{pedge}
S(\omega)\propto (\omega-\omega_0)^{-p},
\end{equation}
where $p=1/2$ and we have defined the edge location $\omega_0 \equiv \langle \omega_{1}\rangle$. Thus, the native SAC edge form with the monotonicity constraint
reproduces the BA form Eq.~(\ref{edgeba}) apart from the logarithmic correction. While this is the asymptotic form close to $\omega_0$ for sufficiently large $N_\omega$,
the spectral profile away from this limit will adapt to the $G(\tau)$ data beyond some distance from $\omega_0$, determined by the statistical errors. 

This edge parametrization can also produce a spectrum with a different exponent $p$ asymptotically by varying the amplitudes as
\begin{equation}\label{eq.varyamp}
 a_i \propto i^c,
\end{equation}
which will result in a spectral function with an edge of the form in Eq.~(\ref{pedge}) for arbitrary $p$ when setting $c=1-2p$. It is useful to let the $i$
dependent amplitudes cross over into the constant form ($c=0$) for $i$ above some sampled value $i_c$, to provide additional flexibility in the way the
asymptotic edge shape adapts to the form favored by the data away from the edge. In Appendix \ref{app:power_adjust} we show an extreme case of such data driven
self-adjustment when $c$ corresponding to a negative $p$ (non-divergent) edge exponent is used in Eq.~(\ref{eq.varyamp}) in a SAC test for a synthetic spectrum
with actual $p>0$; the cross-over point then is pushed to very small $i$ and the imposed wrong asymptotic form becomes irrelevant.

Although in principle the $p=0.5$ form should be suitable for spectra expected to have the BA edge Eq.~(\ref{edgeba}), the logarithmic factor can in practice  
correspond to an apparent effective exponent $p > 0.5$. Moreover, the form is only strictly known for the $2$-spinon contributions, and it is possible that the
exact form is different. We will discuss cases where the optimal $p$ value is significantly larger than $0.5$. 

\subsection{Edge singularity in the QLRO phase}\label{subsec:qlrocase}

In this section we consider $\alpha=3$, which is far inside the QLRO phase. In Fig.~\ref{Fig.a3repsS(k,w)} we show SAC results with the simple edge
constraint described above, i.e., with a lower optimized bound $\omega_0$ imposed on the sampling. We also show results of unconstrained sampling, where
there is spectral weight below the optimal edge. The imposition of the lower edge in this region suppresses configurational entropy and the sampling is
therefore more dominated by spectra with lower $\chi^2$. Though the peak at the edge is sharp, the $G(\tau)$ data quality is not good enough for a divergent
behavior to form because of the entropic pressures suppressing such a peak.

\begin{figure}
\includegraphics[width=8.3cm]{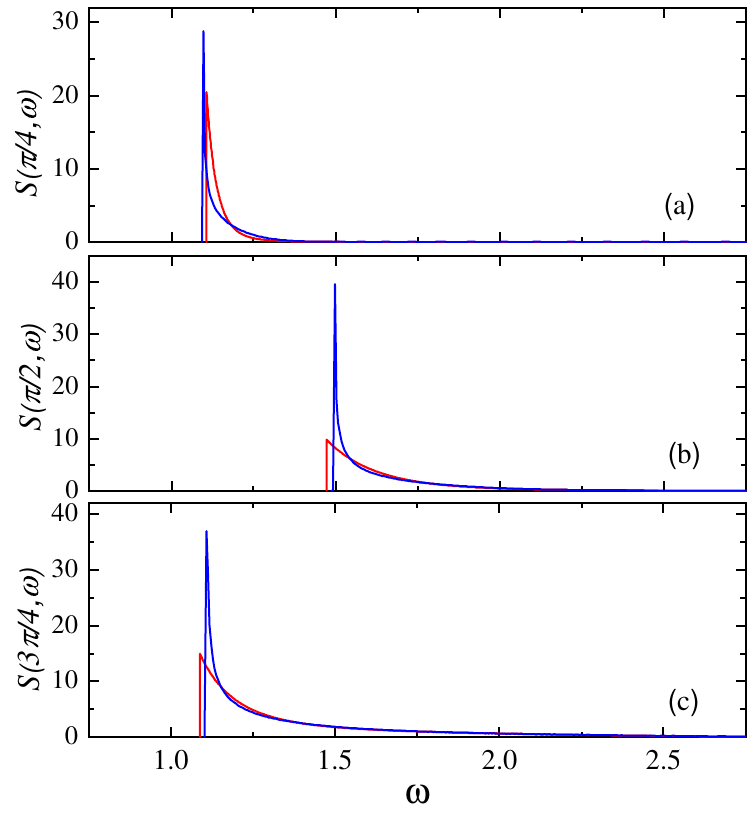}
\caption{Dynamic structure factor of the $L=256$, $\alpha=3$ system at the same momenta as in Fig.~\ref{Fig.a3repsS(k,w)}. Results obtained with the monotonicity
constraint (Fig.~\ref{Fig.sacpara4}), shown in blue, are compared with those of the simple edge from Fig.~\ref{Fig.a3repsS(k,w)}, shown with the red curves.}
\label{Fig.a3edge}
\end{figure}

With the monotonicity constraint illustrated in Fig.~\ref{Fig.sacpara4}, where uniform amplitudes correspond to an edge exponent $p=0.5$, sharp peaks
indeed form, as shown in Fig.~\ref{Fig.a3edge}. The spectra look very similar to those obtained before with this parametrization for the Heisenberg
chain with only $J_1$ interactions \cite{Shao23}. Comparing with the results of the simple edge, it is also notable that the edge itself is located at slightly
higher energy with the monotonicity constraint in Figs.~\ref{Fig.a3edge}(b) and \ref{Fig.a3edge}(c). This behavior can be understood as a compensating effect
of the simple edge when the spectral weight is too small at the location of the edge (which should be very close to the one obtained with the monotonicity
constraint); as a result, some weight is instead pushed below the true edge. In contrast, in Fig.~\ref{Fig.a3edge}(a), even the simple edge produces
a rather sharp peak slightly above the edge location with the monotonicity constraint imposed.

\begin{figure}
\includegraphics[width=8.3cm]{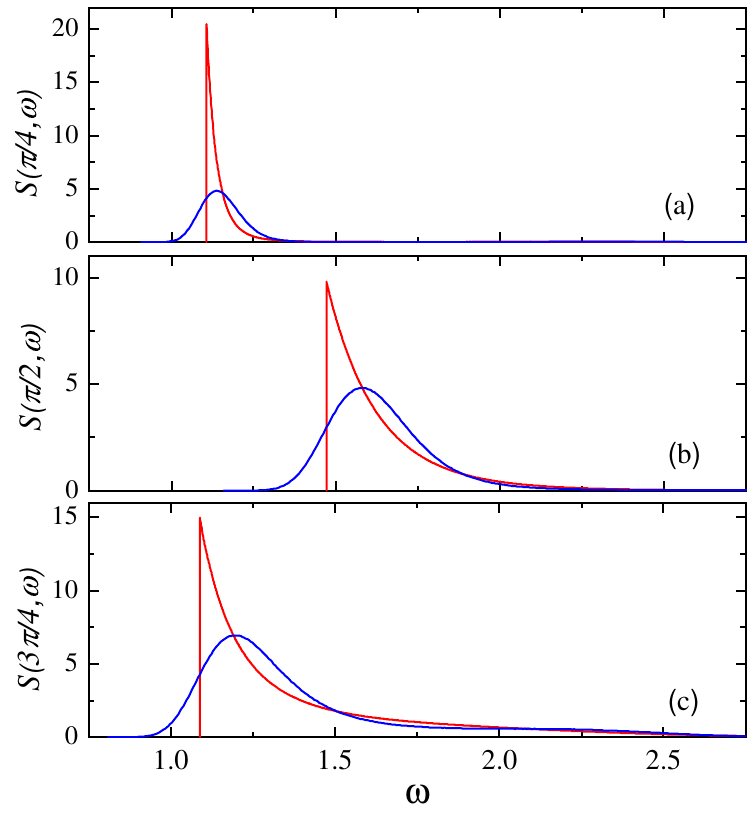}
\caption{Dynamic structure factor of the $L=256$ system at $\alpha=3$ and three different momenta; (a) $k=\pi/4$, (b) $\pi/2$, and (c) $3\pi/4$. The blue curves show
results of unconstrained sampling (with only $\omega$ updates), while the red curves were obtained by imposing a lower bound $\omega_0$ in the sampling. The optimal
bound was found by locating a $\langle \chi^2\rangle$ minimum versus $\omega_0$, in analogy with Fig.~\ref{a0scan}.}
\label{Fig.a3repsS(k,w)}
\end{figure} 

\subsection{Potential edge in the AFM phase}
\label{subsec:afmocase}

\begin{figure}
\includegraphics[width=8.3cm]{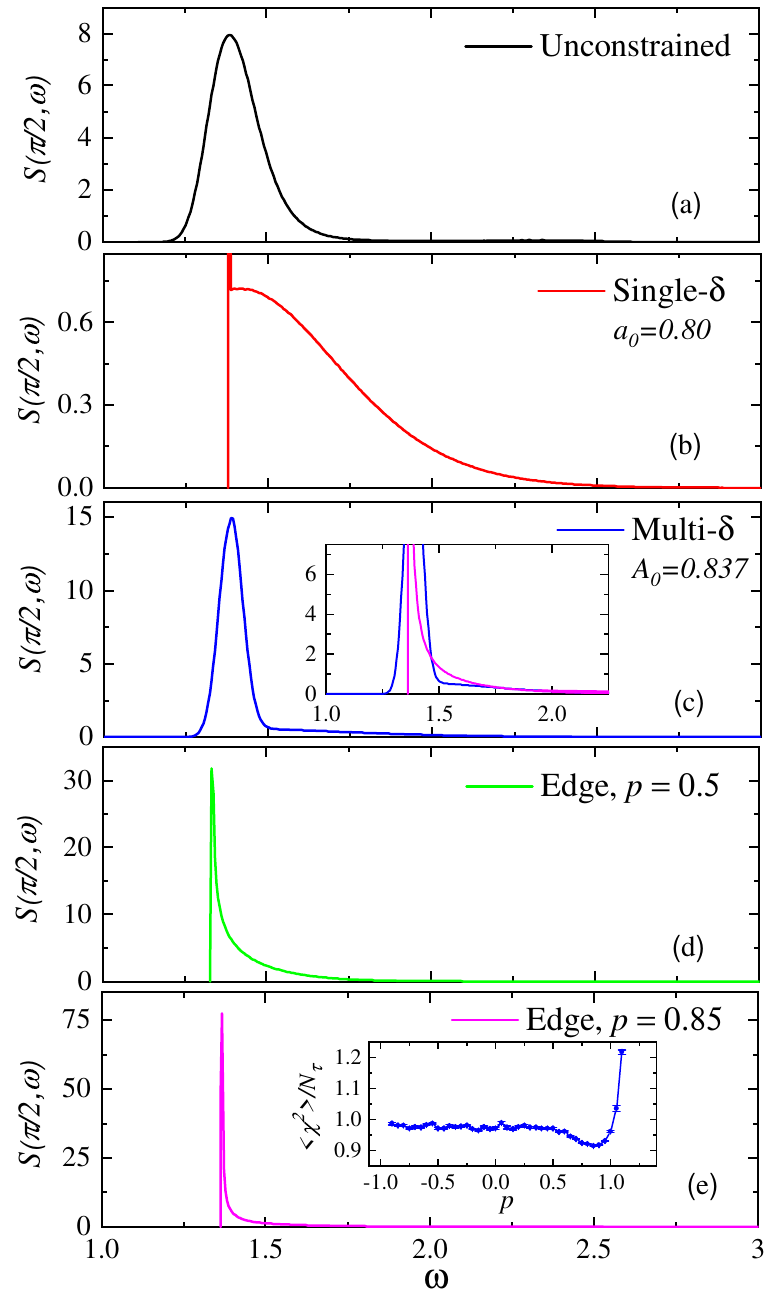}
\caption{Dynamic structure factor for $L=256$, $\alpha=2$, $k=\pi/2$, using five SAC parametrizations; (a) unconstrained, (b) edge with a single
$\delta$-function, (c) broadened multi-$\delta$ peak, (d) edge divergence with the default asymptotic exponent $p=0.5$, and (e) edge with optimized
exponent $p=0.85$. The inset in (e) shows the scan of the mean goodness of fit versus $p$, which delivers the optimal exponent as the location of the minimum.
In the inset of (c), the multi-$\delta$ peak is compared with the edge spectrum with optimal exponent $p=0.85$, same as in (e).}
\label{Fig.JQ3LR_SAC_05pi}
\end{figure}

In Sec.~\ref{subsec:sacNdelta}, we showed that the magnon peak in the AFM phase may be a narrow peak with some broadening, rather than a $\delta$ function.
Given the anomalous nature (nonlinear dispersion relation) of the magnon, an alternative to a broadened, Gaussian-like quasiparticle peak could potentially be an edge
with a power-law divergence. We explore this possibility, using both the native power $p=0.5$ and an optimized power, corresponding to $c\neq 0$ in Eq.~\ref{eq.varyamp}, as determined
via a scan.

In Fig.~\ref{Fig.JQ3LR_SAC_05pi} we show results for $\alpha=2$, using all of the parametrizations discussed above for the case of $k=\pi/2$. The cases of edges with asymptotic exponents $p=0.5$ and $=0.85$ are shown in Figs.~\ref{Fig.JQ3LR_SAC_05pi}(d) and \ref{Fig.JQ3LR_SAC_05pi}(e), respectively. In the
inset of the latter panel we show a scan over the exponent $p$, which exhibits a mean goodness-of-fit minimum at $p=0.85$; the optimal exponent. The
peak here is significantly sharper than what is seen with $p=0.5$ in Fig.~\ref{Fig.JQ3LR_SAC_05pi}(d). We have not found any case for which $p<0$ is optimal;
rather $p>0.5$ is found in all cases.

The inset of Fig.~\ref{Fig.JQ3LR_SAC_05pi}(c) compares the results with the multi-$\delta$ broadened beak and the optimized edge divergence with
$p=0.85$. Here it is obvious that the $p=0.85$ peak is much narrower at the half-height of the multi-$\delta$ peak (corresponding to the maximum value on the
vertical axis). However, at the base of the peak the $p=0.85$ spectrum is broader and can effectively correspond to an overall width similar to that of the
multi-$\delta$ peak.

To distinguish between the $p=0.85$ power-law edge and multi-$\delta$ peak as the best contender for the true spectral form, we can implement a cross validation
procedure within SAC. We carried out tests of this approach in Ref.~\onlinecite{Schumm24}, and in Sec.~\ref{sec:cross_val} we will carry out a more extensive range
of tests. But first, we will perform the simpler test of comparing the dispersion relations extracted using with the single- and multi-$\delta$ parametrizations,
as well as with the power-law edge.

\subsection{Sensitivity of the dispersion relation to the parametrization}\label{sec:dispersion}

\begin{figure}
\includegraphics[width=8.3cm]{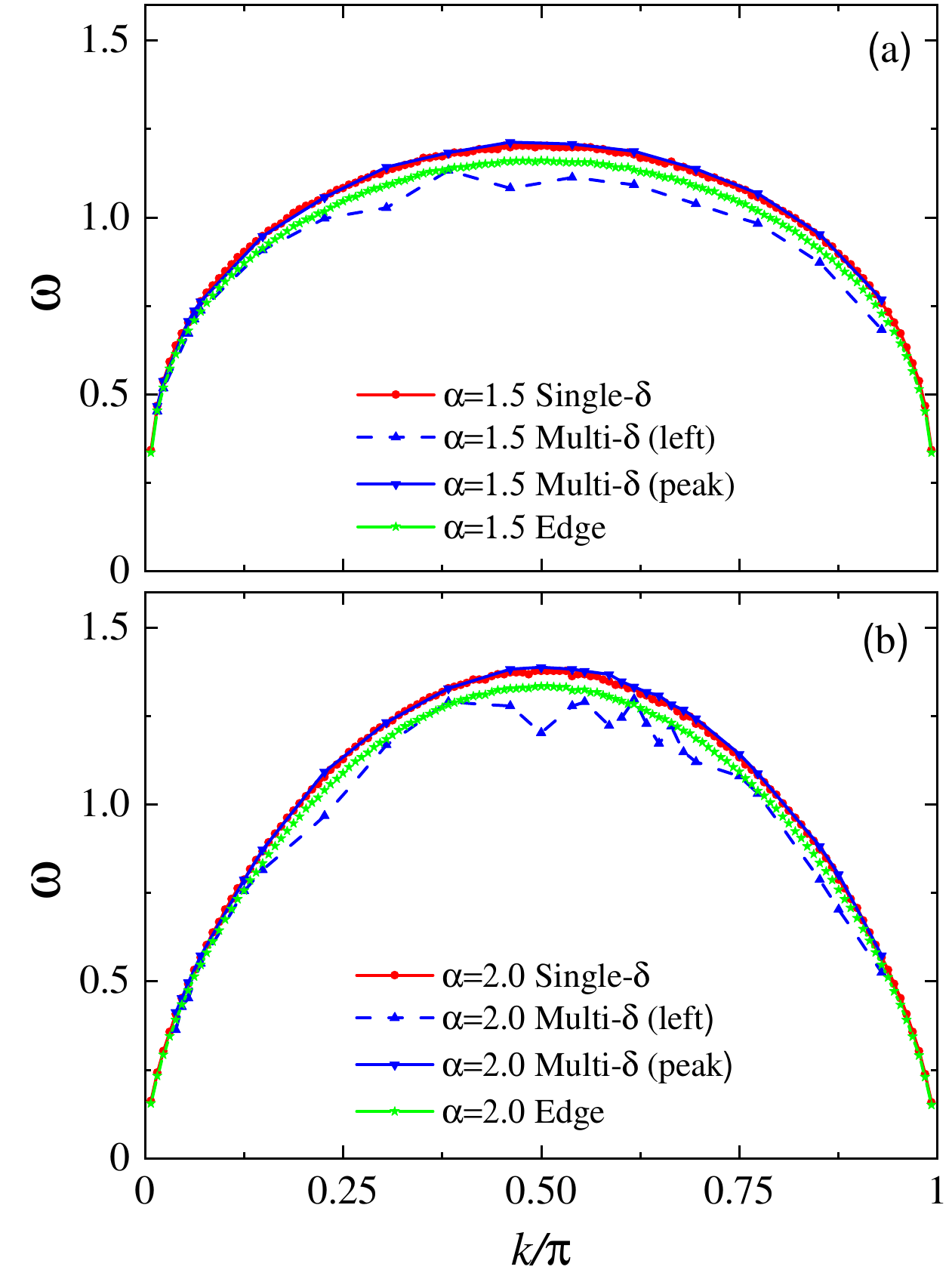}
\caption{Comparison of dispersion relations obtained in SAC with single-$\delta$ peak, multi-$\delta$ broadened peak, and edge singularity with default
exponent $p=0.5$. (a), (b) Show results for the $L=256$ chain with $\alpha=1.5$ and $2.0$, respectively. The red curves showing results for the
single-$\delta$ peak are the same as in Fig.~\ref{Fig.w_vs_k}. In the case of the multi-$\delta$ peak, ``peak'' and ``left'' in the legends refer,
respectively, to the location of the maximum and the point on the low-energy tail where the value is $1\%$ of the peak height. The results labeled
``edge'' represent the $\omega_0$ values obtained with the default edge exponent $p=0.5$.}\label{Fig.dispersion_lp}
\end{figure}

Just as for the single-$\delta$ edge parametrization, we can extract the dispersion relation from the spectra produced by the multi-$\delta$-peak and power-law-edge parametrizations. We show these results in Fig.~\ref{Fig.dispersion_lp} for $\alpha=1.5$ and $2$, along with the previous
single-$\delta$ results for comparison. In the case of the multi-$\delta$ peak, we show results both for the location of the peak and at the lower edge,
which we define as the $\omega$ where the spectral weight is $1\%$ of the peak value. Though not a traditional definition of the dispersion relation
of a broadened quasi-particle peak, this provides another visual representation of the width of the peak. As in Fig.~\ref{width_combine}, the width
is affected by significant statistical uncertainty in the $k$ regions, particularly where this width is the largest. The dispersion extracted by the location of the maximum of the multi-$\delta$
peak coincides very well with that of the single-$\delta$ peak, with small deviations seen only close to $k=\pi/2$ on the scale of
Fig.~\ref{Fig.dispersion_lp}. For the power-law edge, we have used $p=0.5$ for all $k$ here and the peak is then clearly below the peak positions of the other two parametrizations. From Fig.~\ref{Fig.JQ3LR_SAC_05pi}, as well as for other cases that we have tested, the optimal $p$ is larger than $p=1/2$, and when instead using the the optimized divergence exponent,
the edge location is then also much closer to the other values. The differences in the dispersion relations diminish for $k\to 0,\pi$ on account of
the fact that the continuum above the peak becomes very small in these limits. We will address the size dependence of the magnon dispersion relation
in Sec.~\ref{sec:dispersionAFM}, where we extract the exponent governing the nonlinearity for $k\to 0,\pi$.

\begin{figure}
\includegraphics[width=8.3cm]{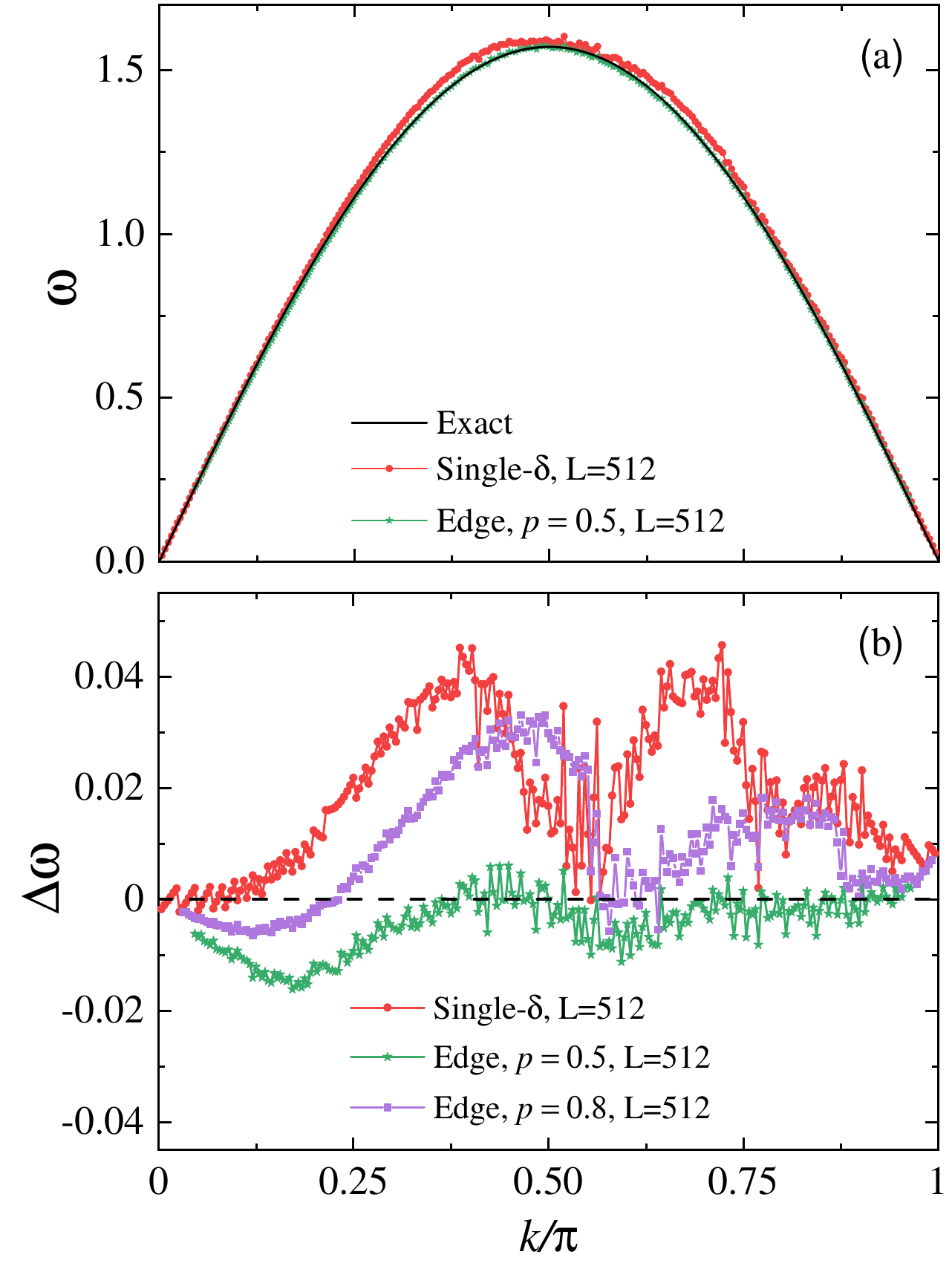}
\caption{(a) Comparison of dispersion relations for the standard $J_1=1$ Heisenberg chain of length $L=512$ obtained with single-$\delta$ magnon
and the power-law continuum edge with default exponent $p=0.5$. The infinite-size exact dispersion relation $\omega=(\pi/2)|\text{sin}(k)|$ is also shown and
coincides closely with the result of the continuum edge parametrization. (b) The deviation of the SAC results from the exact dispersion relation. Here we also
show results with the edge exponent set to $p=0.8$.}
\label{Fig.hchaindisp}
\end{figure}

\begin{figure}
\includegraphics[width=8.3cm]{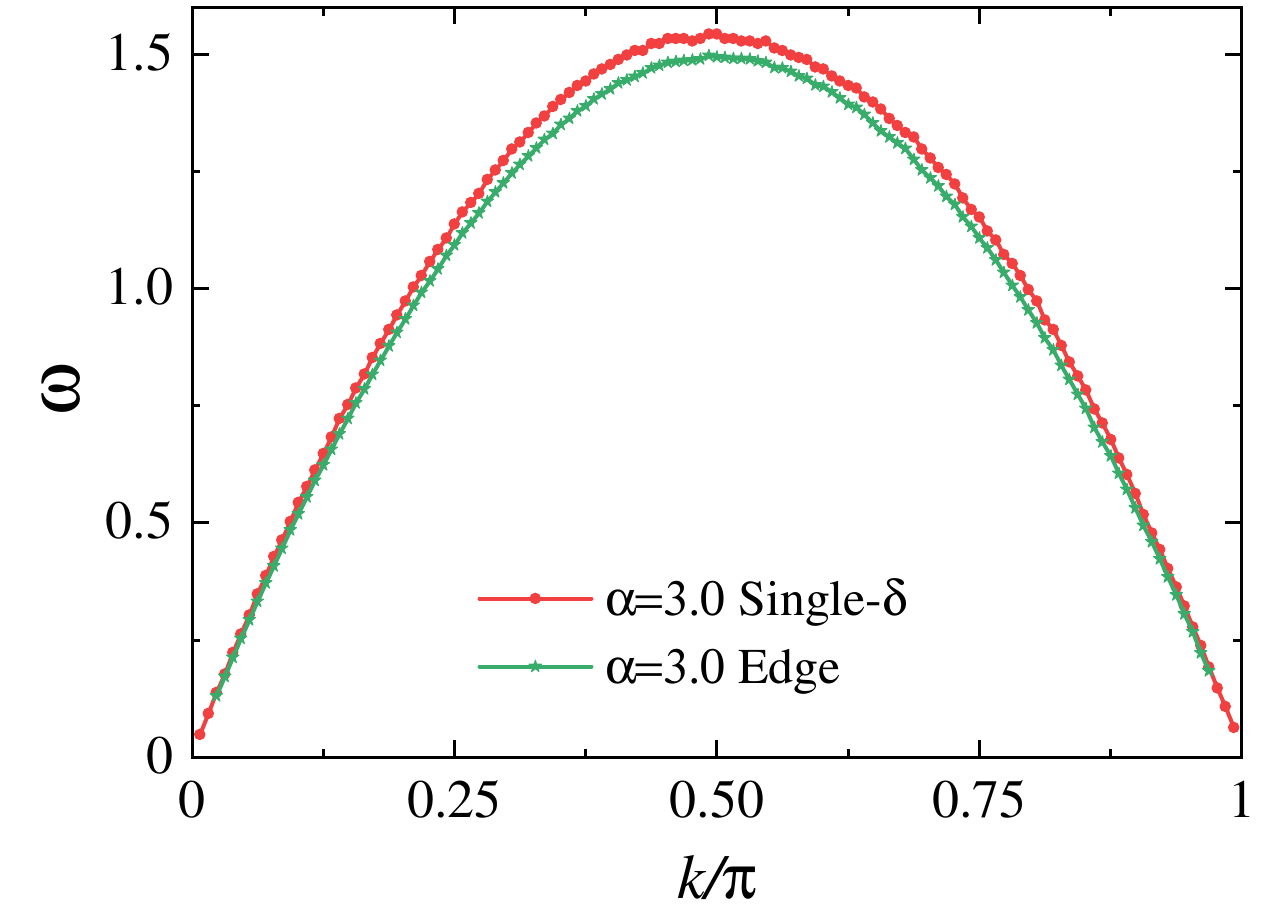}
\caption{Comparison of dispersion relations obtained with single-$\delta$ peak and edge ($p=0.5$) SAC parametrizations for the $L=256$ chain at $\alpha=3.0$.}
\label{Fig.a3disp}
\end{figure}

Turning to the QLRO phase, dispersion relations for the standard $J_1=1$ Heisenberg chain are graphed in Fig.~\ref{Fig.hchaindisp}(a) along
with the exact two-spinon form $\omega_k=(\pi/2)|\text{sin}(k)|$. We compare results for the single-$\delta$ and $p=0.5$ power-law edges.
In Fig.~\ref{Fig.hchaindisp}(b) we show the difference between the exact and SAC dispersions, here including also the edge with $p=0.8$
for further insights into the dependence of the dispersion relation on fine details of the parametrization.

The $p=0.5$ edge produces the best results overall when $k/\pi \agt 0.3$, where the deviations from the exact result are 1$\%$ or below. The
single-$\delta$ edge produces good results for $k$ close to $0$ and $\pi$, and also for a narrow range of $k/\pi$ values close to $0.55$. Away
from these momentum regions, the deviations from the exact dispersion are much larger; up to almost $3\%$. 

Interestingly, for $k/\pi \alt 0.3$, the $p=0.5$ edge performs worse, with the largest deviation around $k/\pi = 0.2$. Here a sharper
edge produces better agreement for small $k$, with $p=0.8$ results shown as an example in Fig.~\ref{Fig.hchaindisp}(b), while for $k/\pi \agt 0.25$
the deviations exceed those for $p=0.5$. By repeating these calculations for smaller systems, we have confirmed that these behaviors are not related to finite-size effects; our results may point to the edge actually being significantly sharper for $k/\pi \alt 0.25$ and not well
modeled with the $p=0.5$ edge parametrization. At $k=\pi/8$, the optimal exponent from the edge SAC parametrization is as large as $p \approx 1.3$, and the excitation
energy is then very close to the result using the single-$\delta$ parametrization and the exact result. A similar case where long-range interactions are included with $\alpha=3.0$ is shown in Fig.~\ref{Fig.a3disp}.

In Ref.~\cite{Caux06}, Caux and Hagemans calculated the four-spinon contributions to the dynamic spin structure factor for the infinite-size Heisenberg
chain. These contributions constitute more than $20\%$ of the total spectral weight and visually exhibit a faster growth upon approaching the edge than the
two-spinon form Eq.~(\ref{edgeba}). However, no quantitative analysis of the singularity was presented. As a follow-up to this work, we are planning
to combine BA and SAC to address the issue of the edge shape in detail \cite{Yang25}. 

\section{Cross validation for model selection}\label{sec:cross_val}

From the tests carried out in the previous sections, one cannot determine which of the SAC parametrizations best describes the spectral function of the
model in the AFM phase. The results of the SAC runs summarized as the five plots in Fig.~\ref{Fig.JQ3LR_SAC_05pi} have acceptable values of $\langle \chi^2\rangle$,
so none of them can be ruled out using a goodness-of-fit criterion alone. The extracted dispersion relations all agree reasonably well with each
other, especially close to $k=0$ and $\pi$, but, to elucidate the full spectral profile, some way to discriminate between different SAC parametrizations
is needed.

A similar problem arises in the field of machine learning when comparing the ability of various models to make predictions based on a set of data. After
training two or more models using the same input data, how can one gauge which model will perform best when new input data are introduced? A common tool used to
tackle this problem is cross validation \cite{mehta_19}. Rather than using the entire data set to train each model, only a subset is used, and the
remaining data are instead used for validation. These validation data are provided as input to the trained models, and scored using some type of loss-function,
typically the same loss function used to train the models initially. The validation score can then be used as a tool to compare the performance of each model.

There is a natural extension of this procedure to SAC. The QMC-generated $G(\tau)$ data are split into independent subsets, which are to be used to calculate averages
and the covariance matrix as described above. Thus, instead of using all $N_B$ bins generated in the SSE simulations to calculate $\bar G(\tau_i)$ and $C_{ij}$, we
split the bins into $K + 1$ mutually exclusive sets of equal length, and these have averages $\bar{G}_n(\tau_i)$ and covariance matrices $C^n_{ij}$, $n=0,\ldots,K$.
For a single cross-validation run, the $n=0$ data will act as the sampling set and are used to calculate the goodness-of-fit $\chi^2 = \chi^2_0$ by Eq.~(\ref{eq.chi2cov})
versus $\Theta$ during an annealing run, while the other $K$ groups are used for cross validation. The loss function in this case is the goodness-of-fit
$\chi^2_\mathrm{val}$ computed with respect to the validation data averaged over each of the $K$ validation sets:
\begin{equation}\label{chi2_val}
\chi^2_{\mathrm{val}} = \frac{1}{K} \sum_{n=1}^K \sum_{i=1} \frac{1}{\sigma_n(\tau_i)^2} \left[ G_S(\tau_i) - \bar{G}_n(\tau_i)\right]^2,
\end{equation}
where $S$ refers to the instances of the spectral function generated in the sampling run according to $\chi^2_0$. 

\begin{figure}
\includegraphics[width=8.0cm]{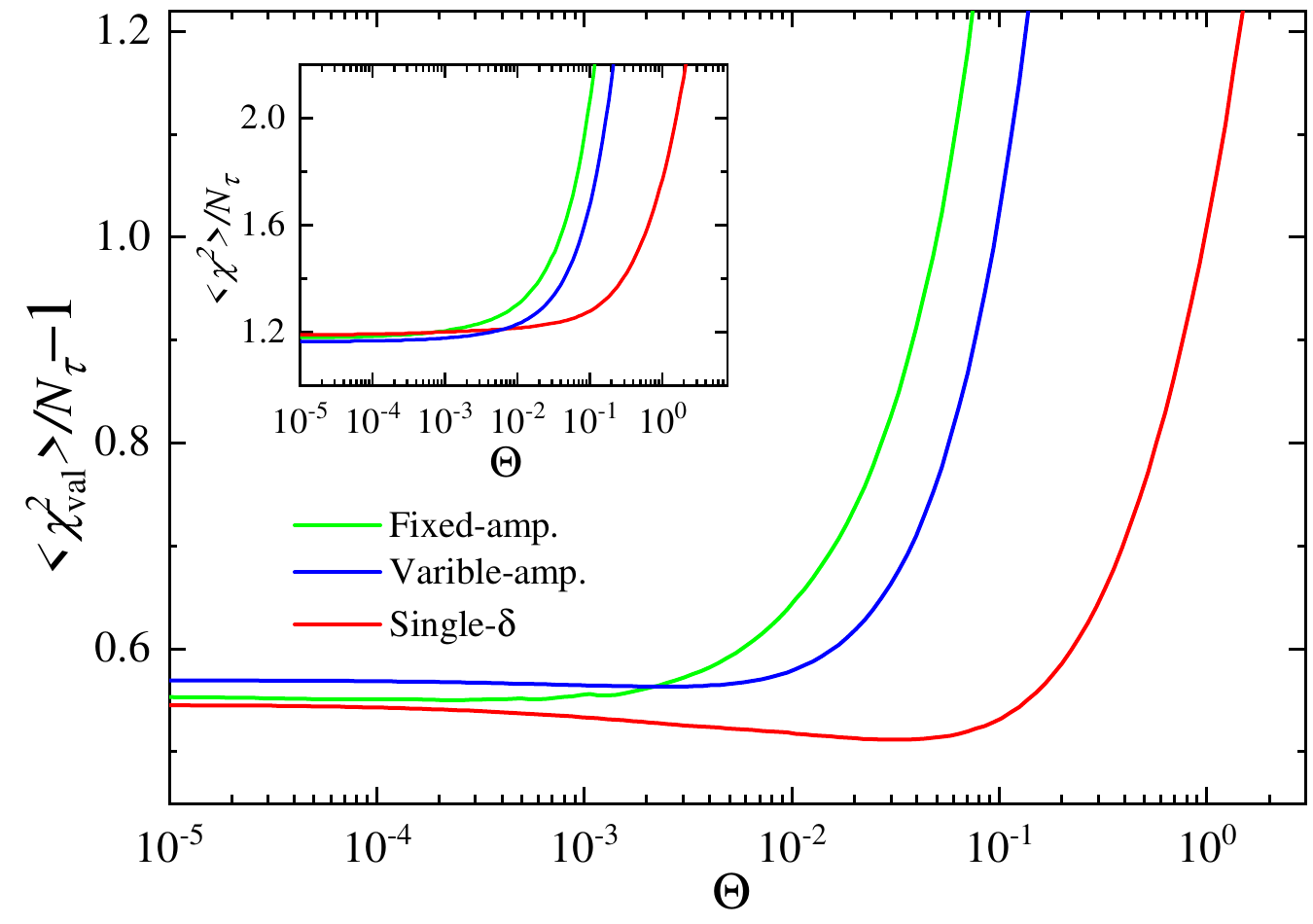}
\caption{Cross validation results for the $L = 256$ chain with $\alpha = 2$, $k = \pi/128$. The average of the validation $\chi^2$ value, Eq.~(\ref{chi2_val}),
normalized by the number of $\tau$ points and with the constant $1$ offset removed, is shown in the main graph, The average over the conventional sampling
$\chi^2$ value during the anneal runs is shown in the inset. For this test, $K+1 = 21$ total data sets were used, each containing $\sim 260$ QMC bins.}
\label{Fig.a2ampcv}
\end{figure}

Cross validation was first combined with SAC in Ref.~\onlinecite{Efremkin21} and in a recent study \cite{Schumm24} we further
developed the approach and emphasized its use as a model selection tool. Because of the different entropic pressures affecting the average spectrum generated
with different parametrization, one can regard parametrizations as models that are optimized (trained) in the SAC procedure to fit (but not overfit) the QMC data.
We have already applied cross validation to some preliminary results for the model considered here \cite{Schumm24}, but in this section we present calculations
with further improved $G(\tau)$ data and consider a wider set of SAC models, including the multi-$\delta$ peak parametrization and optimal $p$ edge.

While $\langle \chi^2(\Theta)\rangle$ decreases monotonically in an SAC annealing run with slowly lowered sampling temperature $\Theta$ (ideally staying in equilibrium
for all $\Theta$), $\langle \chi^2_{\mathrm{val}}(\Theta)\rangle$ is expected to exhibit a minimum and then increase below some low $\Theta$ value as a result of overfitting
to the training data. To take full advantage of the information contained in the data, we rotate which of the $K+1$ data sets is used for sampling (with the remaining
$K$ sets now used for validation). The final validation result is the average over all of these rotations; a single curve of $\langle \chi^2_{\mathrm{val}}\rangle$
versus $\Theta$ for each parametrization. Here we take $K$ to be of the order $10$. As explained in Refs.~\onlinecite{Efremkin21} and \onlinecite{Schumm24},
$\langle \chi^2_{\mathrm{val}}\rangle$ contains a trivial constant $N_\tau$. We remove this constant and present results for both $\langle \chi^2(\Theta)\rangle/N_\tau$
(which is also averaged over the rotations) and $\langle \chi^2_{\mathrm{val}}(\Theta)\rangle/N_\tau-1$.

We begin by showing in Fig.~\ref{Fig.a2ampcv} cross-validation results corresponding to the spectral functions in the AFM phase in Fig.~\ref{Fig.a2repsS(k,w)}(a)  ($\alpha = 2$),
comparing the single-$\delta$ magnon peak and the two unconstrained parametrizations. Here the single-$\delta$ peak clearly produces the lowest
$\langle \chi^2_{\mathrm{val}}\rangle$ value for all $\Theta$, and there is also a clear minimum. For the unconstrained parametrizations, either there is only a very shallow
minimum in $\langle \chi^2_{\mathrm{val}}\rangle$ (when amplitude sampling is included) or the minimum is in the $\Theta \to 0$ limit (when only frequencies are sampled).
This test solidifies the parametrization with the single-$\delta$ magnon mode as the best out of these three models. Note that this conclusion cannot
be drawn based on the sampling goodness-of-fit, which, as shown in the inset of Fig.~\ref{Fig.a2ampcv}, is acceptable for all three parametrizations, but not the
lowest for the single-$\delta$ peak in the limit of $\Theta \to 0$.

\begin{figure}
\includegraphics[width=8.3cm]{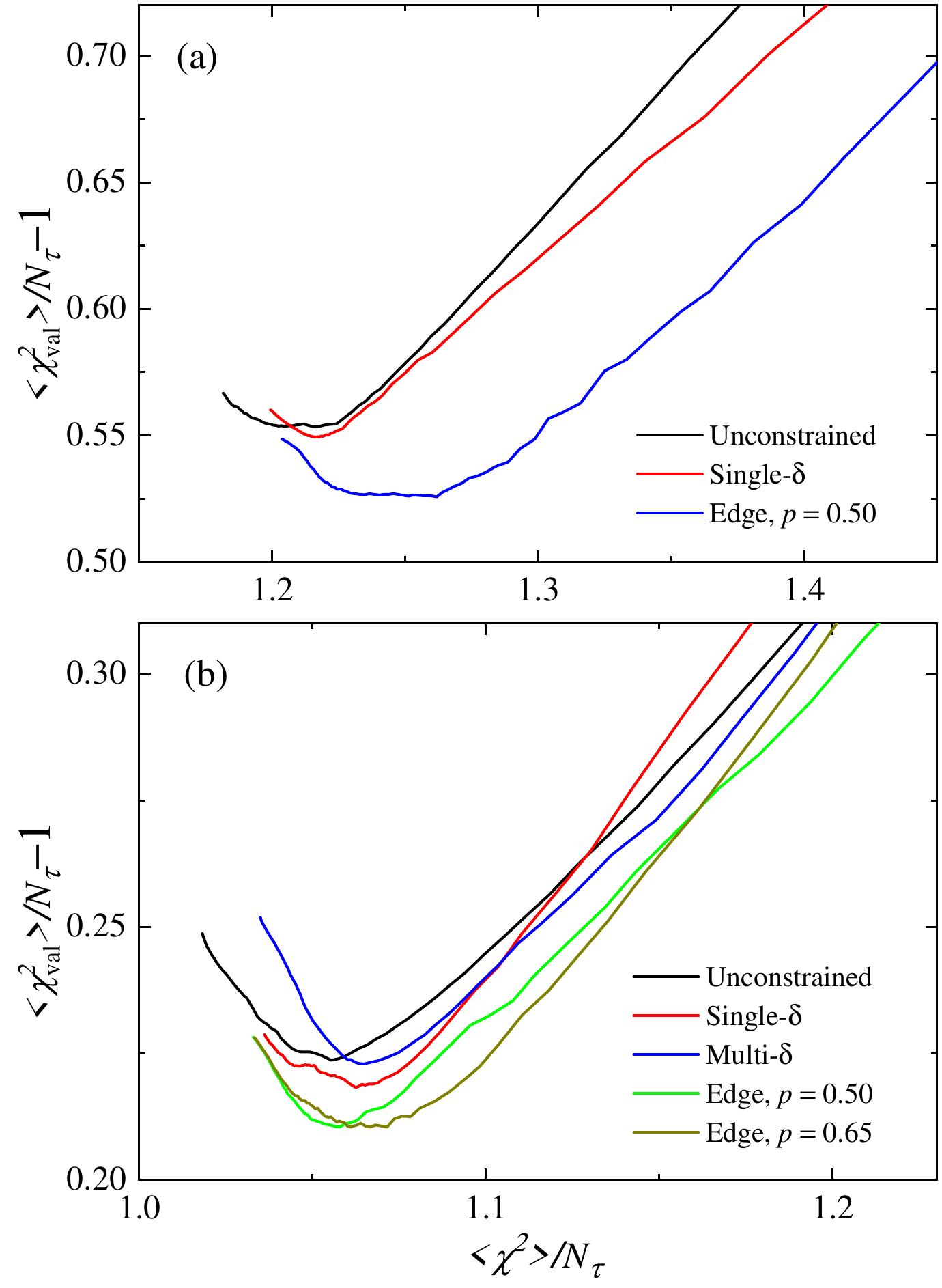}
\caption{Cross-validation results in the QLRO phase for $k=\pi/2$. Results for the $J_1=1$ Heisenberg chain and the long-range chain with
$\alpha=3$ are shown in (a) and (b) respectively. The normalized average validation goodness-of-fit $\langle \chi^2_{\rm val}\rangle$ (with the offset
removed) is here graphed vs the conventional normalized sampling average $\langle \chi^2\rangle$. In (a) the result of unconstrained SAC (including
amplitude updates) is compared with those of the single-$\delta$ magnon and $p=0.5$ power-law edge, while in (b) also results of an optimized $p=0.65$
edge and a multi-$\delta$ broadened magnon are included.}
\label{Fig.QLRO_CV}
\end{figure}

In the QLRO phase, we would expect the power-law edge to be the best parametrization, ideally with the exponent $p$ optimized. In Fig.~\ref{Fig.QLRO_CV}(a), we consider the plain $J_1=1$ Heisenberg chain at $k=\pi/2$, using only the default exponent $p=0.5$, which produced an excellent agreement with the exact $\omega_k$ in Fig.~\ref{Fig.hchaindisp}. Indeed, this edge parametrization validates far better than the unconstrained and single-$\delta$ peak parametrizations. Here, as an alternative to graphing results versus
$\Theta$ as in Fig.~\ref{Fig.a2ampcv}, we have graphed $\langle \chi^2_{\mathrm{val}}(\Theta)\rangle/N_\tau-1$ versus $\langle \chi^2(\Theta)\rangle/N_\tau$, which allows
direct comparisons of the different models at the same value of the conventional goodness of fit. 

\begin{figure}
\includegraphics[width=8.5cm]{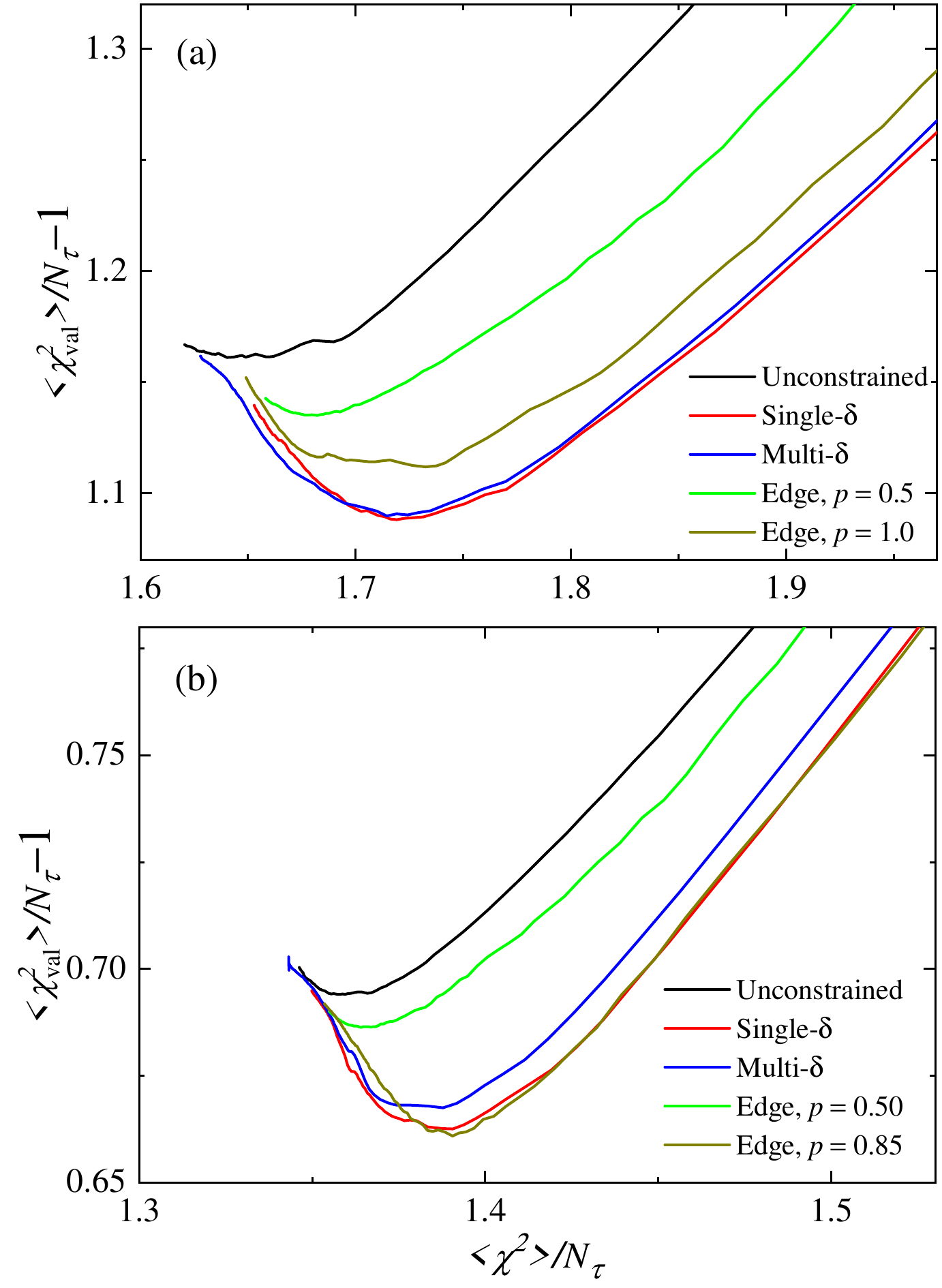}
\caption{Cross validation results graphed in the same way as in Fig.~\ref{Fig.QLRO_CV} for the $L=256$ chain with $\alpha = 1.5$
in (a) and $\alpha=2$ in (b), all for $k=\pi/2$. Five different parametrizations are compared, including the power-law edge with optimized exponent $p=1$
and (a) and $p=0.85$ in (b).}
\label{Fig.LRa15_CV}
\end{figure}

In Fig.~\ref{Fig.QLRO_CV}(b), we consider the long-range model with $\alpha = 3$, and now include results obtained with the optimized edge exponent $p=0.65$ and the multi-$\delta$ peak. The optimized edge is now validated
as the best case, though it is only marginally better than the default $p=0.5$ edge.

The above results represent cases where cross-validation can single out the best model out of a set with rather high confidence, in Fig.~\ref{Fig.a2ampcv} likely
because the momentum is close to $0$ so that the expected magnon peak in the AFM phase must be very narrow. Similarly, in Fig.~\ref{Fig.QLRO_CV} analogously
the systems are deep inside the QLRO phase where the power-law edge should clearly be better than the other representations. Although the results are expected,
it is still very gratifying to see that the method indeed is able to distinguish the different cases even though all the models work well in terms of the
conventional goodness of fit.

Cross validation becomes more challenging in cases where the true spectrum is likely not completely reproducible by any of the models. For $\alpha$ inside the AFM
phase, but close to the AFM--QLRO transition, the spectrum can be expected to exhibit a continuum closely mimicking a power-law edge, but with eventually a magnon peak
instead of a power-law divergence very close to the edge. In Fig.~\ref{Fig.LRa15_CV} we show results for $k=\pi/2$ at $\alpha=1.5$ and $2.0$, using the same
parametrizations as before and including the power-law edge with both default and optimized exponents $p$. At $\alpha=1.5$, Fig.~\ref{Fig.LRa15_CV}(a), the optimized
edge with $p=1.0$ performs better than the default $p=0.5$ edge, but neither of them is close to the best-validated case of the single-$\delta$ peak, which, in turn,
is only marginally better than the multi-$\delta$ peak case. Closer to the AFM--QLRO transition, at $\alpha=2$ in Fig.~\ref{Fig.LRa15_CV}(b), the multi-$\delta$ peak
performs worse relative to the single-$\delta$ peak, which now competes very closely with the power-law edge with optimized $p$. Here the optimized edge, which
has a large exponent $p=0.85$, likely performs well because it makes a successful (in a statistical sense) approximation to a small sharp magnon peak superimposed
on an almost divergent continuum resembling that in the nearby QLRO phase at $\alpha > \alpha_c \approx 2.23$.

While there clearly are some limitations to cross-validation as a model selection tool, our results show that it works well in cases where the spectral function
does not mix different features that are difficult to capture in combination by the SAC method (or any other analytic continuation method), ultimately because of insufficient
data quality. The set of parametrizations considered is also clearly of importance. When two or more parametrizations perform comparably, ideally further refined
parametrizations and constraints can be introduced with the goal of further improving the validation goodness-of-fit. In the case of $\alpha$ close  
to $\alpha_c$, a $\delta$-function edge could in principle be combined with the monotonicity constrained continuum, Fig.~\ref{Fig.sacpara4}, where a minimum distance
is imposed between the first two $\delta$-functions, which quenches the divergence of the continuum \cite{Shao23}. We have not yet explored this more complicated
parametrization.

\section{Dispersion relation and dynamic exponent in the AFM phase}\label{sec:dispersionAFM}

\begin{figure}
\includegraphics[width=8.5cm]{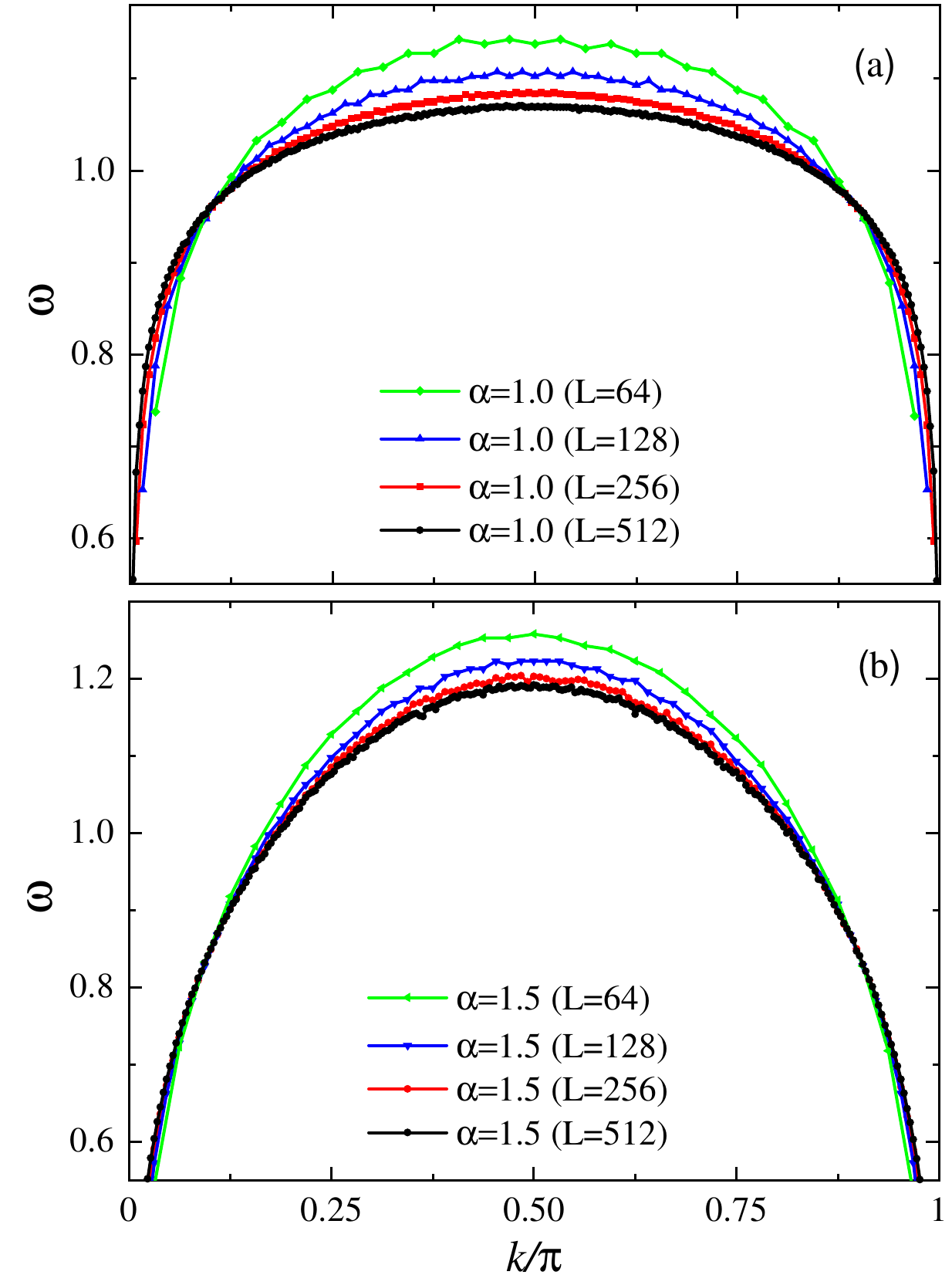}
\caption{Size dependent dispersion relation $\omega_k$ for $\alpha=1.0$ (a) and $1.5$ (b), in both cases with chain lengths $L=64$, $128$, $256$, and $512$.
The points correspond to the position $\omega_k$ of the macroscopic $\delta$ function in the single-$\delta$ edge SAC.}
\label{Fig.dispsizes}
\end{figure}

\begin{figure}
\includegraphics[width=8.5cm]{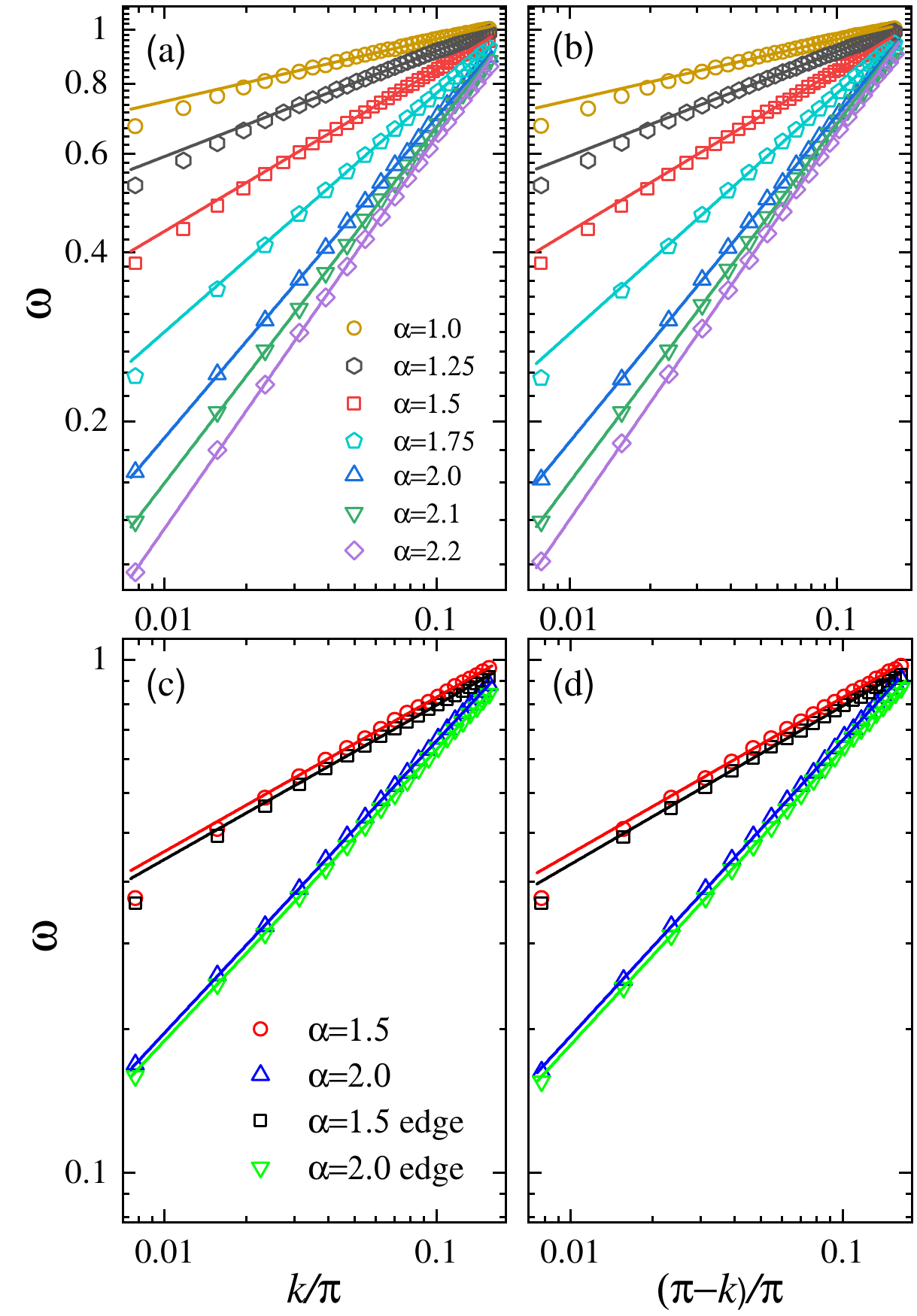}
\caption{Dispersion relation close to (a) $k=0$ and (b) $k=\pi$ for several values of $\alpha$, with the results obtained for chain length
$L=512$ for $\alpha \le 1.5$ and $L=256$ for $\alpha > 1.5$. Log-log scales are used to observe the power-law form $\omega_k \sim k^z$ in (a)
and $\omega_k \sim (\pi-k)^z$ in (b). The lines are fits that for $\alpha < 2$ exclude some of the the points closest to $0$ and $\pi$,
where finite-size effects are still present. For given $\alpha$, the lines have the same slope in (a) and (b).}
\label{Fig.powerfit}
\end{figure}

Based on the cross-validation tests in Sec.~\ref{sec:cross_val}, we can conclude that the single-peak SAC performs best in the AFM phase, as expected, though
in principle there could be some small broadening that we cannot resolve. In the limits $k \to 0$ and $k \to \pi$, any broadening and continuum should vanish and
we can reliably extract the dynamic exponent $z$ from the peak location, $\omega_k\propto k^z$. We already showed results for the full dispersion relation for different
values of $\alpha$ in Fig.~\ref{Fig.w_vs_k}. We used the chain length $L=256$ for all system sizes, but finite-size effects are large, especially when $\alpha$ is below
$1.5$. The finite-size evolution of the dispersion relation must be investigated before extracting the dynamic exponent.

Figures \ref{Fig.dispsizes}(a) and \ref{Fig.dispsizes}(b) show dispersion relations for several system sizes at $\alpha=1.0$ and $1.5$, respectively. Judging from
these results, at $\alpha=1.5$ there should only be small remaining finite-size effects with $L=512$, while there are still finite-size effects left at $\alpha=1$.
In Fig.~\ref{Fig.powerfit} we show results for the largest system size studied for several values of $\alpha$, using log-log scales for data close to $0$ in
Fig.~\ref{Fig.powerfit}(a) and close to $\pi$ in Fig.~\ref{Fig.powerfit}(b) (in the latter case graphing versus $\pi-k$). For $\alpha \ge 2$, clean power law
scaling is observed all the way from the smallest value $2\pi/L$ of $k$ or $\pi-k$, while for smaller $\alpha$ there are clear deviations that we ascribe to remaining
finite-size effects. We have carried out fits with those points left out, finding the same dynamic exponent for $k$ close to $0$ and $\pi$, as expected. The results are
graphed versus $\alpha$ in Fig.~\ref{Fig.powerfit}.

Given that the nature of the low-energy edge likely is influenced by the power-law divergent continuum when $\alpha$ is close to the transition point $\alpha_c$ to
the QLRO phase, we also have extracted the dynamic exponent using the $p=0.5$ edge form. As already seen in Fig.~\ref{Fig.dispersion_lp}, the dispersion relations close
to $k=0$ and $\pi$ are not sensitive to the form of the edge used. In Figs.~\ref{Fig.powerfit}(c) and \ref{Fig.powerfit}(d) we compare fits to the appropriate
low-energy segments of the dispersion relation to results from both the single-$\delta$ and $p=0$ edges. While there is a small overall shift of the energies,
the dynamic exponent is essentially the same.

\begin{figure}
\includegraphics[width=8.5cm]{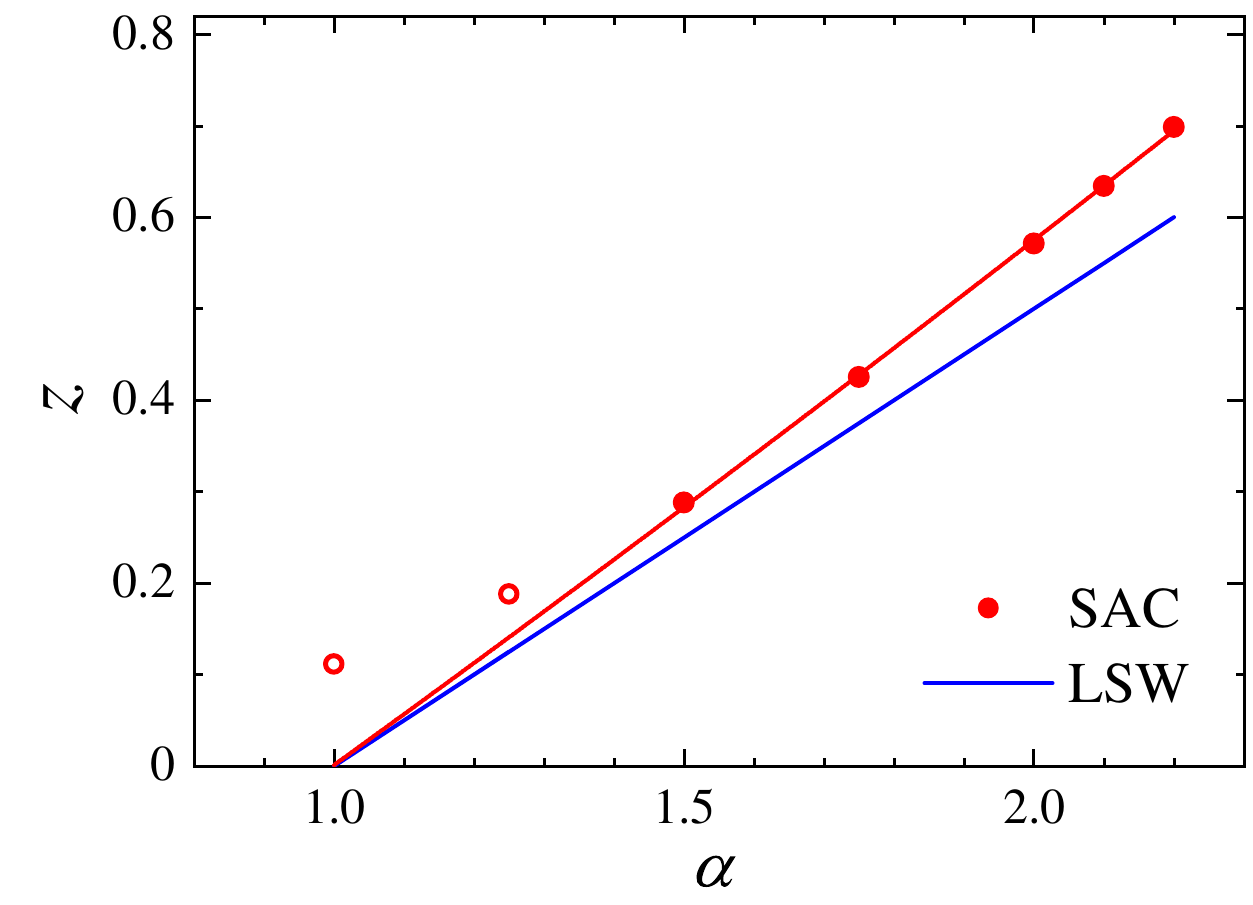}
\caption{Dynamic exponent versus $\alpha$ obtained from the fits in Fig.~\ref{Fig.powerfit} compared with the SWT result (blue line). The solid and open red circles correspond to well converged and likely not fully converged results, respectively. The
red line shows a second-order polynomial fit to the converged SAC results, assuming $\alpha_s=1$. The quadratic correction is very small, and leaving it out
also leads to an acceptable fit.}
\label{Fig.exponents}
\end{figure}

In Fig.~\ref{Fig.exponents} we compare $z(\alpha)$ from the single-$\delta$ edge with SWT calculations, first carried out by Yusuf et al.~\cite{Yusuf04} and
adapted in Appendix \ref{appsec:swa1a2} to the exact form of the interaction in our model. If the SAFM--AFM transition takes place at $\alpha=1$ as in linear
SWT, $z$ should approach $0$ at that point. While our results for $z(\alpha)$ are consistently slightly above the SWT result, the behavior is still consistent
with $z \to 0$ as $\alpha \to 1$ if we disregard the likely not fully size-converged results for $\alpha=1$ and $1.25$. A fit of our results to a second-order
polynomial is also shown in Fig.~\ref{Fig.exponents}. The quadratic correction is very small and a linear form, as in SWT, also provides an acceptable
fit to the data. For $\alpha$ close to $\alpha_c \approx 2.23$, our value of $z$ is close to previous estimates $z(\alpha_c)\approx 3/4$ based on other methods
\cite{Laflorencie05,Sandvik10}.

\section{Summary and discussion}\label{sec:summary}

We have investigated the dynamic spin structure factor $S(k,\omega)$ in the ground state of the spin-1/2 Heisenberg chain with long-range sign-alternating
(bipartite), power-law decaying interactions. Using the SAC method with high-quality imaginary-time correlations computed using SSE-QMC simulations, we have applied
different types of constraints in order to resolve the sharp feature expected at the lowest excitation energy $\omega_k$. We further used cross-validation to
single out the optimal constraint (shape of the edge) in the AFM and QLRO phases.

Sufficiently deep inside the AFM phase, our results support a $\delta$-function edge corresponding to a single-magnon excitation. Although we cannot strictly rule
out that the peak is broadened by interactions, with one of our representations indeed producing a broadened quasi-particle peak, cross-validation does not support
broadening below a sharp edge. There is, however, a significant continuum above the edge, which vanishes only in the limits $k\to 0$ and $k \to \pi$.

Close to the quantum phase transition into the QLRO phase, the continuum starts to develop the features of a power-law divergent continuum that is expected inside the
QLRO phase. In the AFM phase, a likely interpretation of our results is that the divergence of the continuum is quenched, and instead a $\delta$ function
related to the magnon in the long-range-ordered system emerges at the otherwise nonsingular edge.

We investigated the relative weight of the single-magnon contribution versus the momentum at different points in the AFM phase. However, because the combination of
the near-critical form of the continuum and the magnon peak is not fully captured by parametrization used here close to the QLRO phase, the results for the relative
magnon amplitude versus $k$ in Fig.~\ref{width_combine}(a) are likely reliable only deep inside the AFM phase (while close to the QLRO phase the weight $a_0$ also
contains contributions from a very narrow continuum close to the edge). It would be interesting to repeat the calculations with a SAC parametrization properly combining
the near-divergent continuum and the magnon peak, which, however, is beyond the scope of this survey.

In the QLRO phase, our parametrization of the divergent edge produces results in close agreement with the expected form, as has been previously shown for
the standard Heisenberg chain ($J_1$ interactions only) \cite{Shao23} for $k/\pi > 0.25$. For smaller $k$, we have found signs of a faster divergence at
the edge than the conventional two-spinon form. We plan to investigate the Heisenberg chain in more detail in the future, using a combination of finite-size
numerical BA and QMC-SAC calculations \cite{Yang25}.

Despite the potential deficits in modeling the detailed form of the spectral edge in the near-critical part of the AFM phase, we have shown that the dispersion
relation in the low-energy limits $k \to 0$ and $k \to \pi$ is insensitive to the exact form of the parametrization used. The nonlinear form of the dispersion
relation corresponds to a dynamic exponent $z<1$, which was known from previous SWT calculations. For $\alpha$ close to $1$, large finite-size effects on the dispersion
relation makes it impossible to extract a reliable value of $z$. For $\alpha=1.5$ and higher, our results are well converged and consistently give values of $z$
somewhat larger than the SWT results, though still being consistent with $z \to 0$ as $\alpha \to 1$. In SWT, $z(\alpha)$ is linear in $\alpha$ and our results
are consistent with such a form as well, with a slightly larger coefficient and possibly a small quadratic correction. 

Our study demonstrates the versatility of the SAC approach with constraints to model sharp spectral features, as well as the use of cross validation as a SAC model
(i.e., constrained parametrization) selection tool. An interesting aspect of cross validation
exemplified by this study is that two (or possible more) SAC models can be essentially equally validated, which then calls for the construction of yet a better
parametrization that in some ways combines features from both. In the present case, our results in the AFM phase close to the QLRO boundary presented this behavior
and suggested a new parametrization as discussed above. While one can of course never guarantee that the optimal model has been found, our approach can at
least point the way to improved models and spectral functions exhibiting details that would be impossible to obtain with conventional numerical
analytic continuation tools.

\begin{acknowledgments}
This research was supported by the Simons Foundation under Grant No.~511064. The numerical calculations were carried out on the Shared Computing Cluster
managed by Boston University’s Research Computing Services.
\end{acknowledgments}

\appendix

\section{Spin-wave calculations}\label{appsec:swa1a2}

Though the SWT calculations for the model are standard \cite{Yusuf04,Laflorencie05}, we briefly review its implementation for the model in the form of
Eq.~(\ref{jrdef}). The main purpose here is to investigate the size dependence of the dispersion relation, which shows similarities with our results from
QMC-SAC in its slow convergence deep inside the AFM phase, close to $\alpha=1$, and fast convergence closer to the QLRO transition at larger $\alpha$.

We first divide the chain into two sublattices and represent the spin operators accordingly in terms of two types of bosons: $a$ and $b$ bosons on sublattice
$A$ and $B$, respectively. Up to the order of $1/S$, where $S$ is the size of the spin, the Holstein-Primakoff transformation for the spin operators can be
written as
\begin{equation}
\begin{aligned}
 S_i^z= &S-a_i^\dagger a_i , \quad S_i^-=\sqrt{2S}a_i^\dagger[1-a_i^\dagger a_i/(2S)]^{1/2}\simeq\sqrt{2S}a_i^\dagger; \\
 &i\in \text{odd},\\
 S_j^z= &b_j^\dagger b_j-S , \quad S_j^-=\sqrt{2S}[1-b_j^\dagger b_j/(2S)]^{1/2}b_j\simeq\sqrt{2S}b_j. \\
  &j\in \text{even}.\\
\end{aligned}
\end{equation}
After the standard steps of (1) Rewriting the Hamiltonian in terms of the above $a$ and $b$ bosons, (2) Fourier transforming to momentum space, and (3)
carrying out the  Bogoliubov transformation to diagonalize the Hamiltonian, we obtain the final simplified expression
\begin{equation}
 H=\text{const.}+GS\sum_k\omega_k(c_k^\dagger c_k+d_k^\dagger d_k),
\end{equation}
where $c_k$ and $d_k$ are free boson operators and 
\begin{equation}
 \omega_k=\sqrt{[\gamma-f(k)]^2-[g(k)]^2},
\end{equation}
where 
\begin{equation}
\begin{aligned}
\gamma=2\sum_{r=1}^{L/4}\frac{1}{(2r-1)^\alpha},\\
f(k)=2\sum_{r=1}^{L/4}\frac{\text{cos}(2rk)-1}{(2r)^\alpha},\\
g(k)=2\sum_{r=1}^{L/4}\frac{\text{cos}[(2r-1)k]}{(2r-1)^\alpha}.
\end{aligned}
\end{equation}
The only difference between $\omega_k$ and the results in Refs.~\cite{Yusuf04,Laflorencie05} is the summation limit, where their long-range interaction
in the previous works extends to $L$ rather than $L/2$ used here. To compare with the QMC-SAC results, we divide all energies by $GS$, where $G$ is the common
factor used to normalize the interaction in Eq.~(\ref{jrdef}).

\begin{figure}
\includegraphics[width=8.3cm]{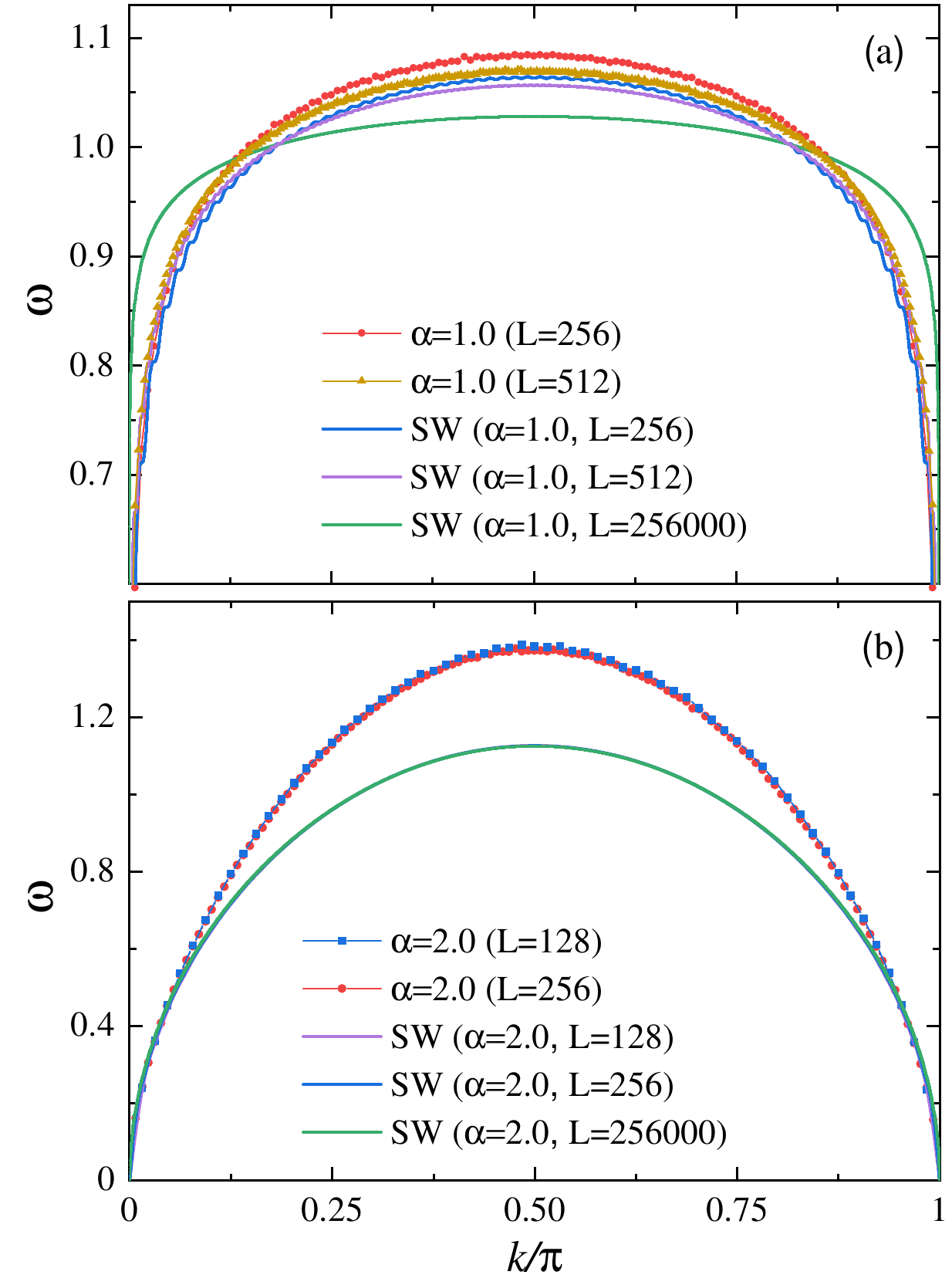}
\caption{Comparisons of the dispersion relation from the single-peak SAC and the linear SWT expressions using the same chain lengths $L=256$
and $512$ in the QMC-SAC calculations and linear SWT. In the case of SWT, the result for $L=256000$ are also shown. In (a) the case of $\alpha=1$
exhibits slow convergence while the results for $\alpha=2$ in (b) converge rapidly. Here all SWT results fall essentially on the same curve.}
\label{Fig.swa1a2}
\end{figure}
  
In Fig.~\ref{Fig.swa1a2} we show illustrative results for the finite-size dispersion relation the long-range exponent $\alpha=1$ and $\alpha=2$, comparing
the results obtained with SAC with the single-$\delta$ edge. For $\alpha=1$ in Fig.~\ref{Fig.swa1a2}(a), the system is deep in the AFM phase and the
both methods show significant finite-size effects. Within SWT, $\alpha=1=\alpha_s$ is the point at which the SAFM phase with flat dispersion is entered,
but even for $L=256000$ there is still significant dependence on the SWT dispersion. In contrast, for $\alpha=2$ in Fig.~\ref{Fig.swa1a2}(b), the
results are well converged already for $L=256$. The QMC-SAC results also suggest fast convergence at $\alpha=2$, while for $\alpha=1$ the results
are likely not completely converged for the largest system size studied, $L=512$. In fact, the $L=256$ and $L=512$ dispersion relations are
close to each other for a substantial range of $k$ values close to $0$ and $\pi$, as we have discussed in more detail in Sec.~\ref{sec:dispersionAFM},
and it is possible to extract the dynamic exponent $z > 0$ here.

\section{Multi-$\delta$ peak optimization}
\label{app:multipeak}

\subsection{Scanning procedure}\label{appsubsec:mpsac}

As a faster alternative to a 2D scan to locate the optimal $(A_0,N_p)$ in the multi-$\delta$ peak parametrization (Fig.~\ref{Fig.sacpara3}),
Ref.~\cite{Shao23} discussed a two-line scan procedure that seems to work well in test cases though not always being fully optimal in locating the
true $\langle \chi^2\rangle$ minimum. We here apply the two-line scan with our real QMC data, incorporating a further improvement in the determination
of $A_0$. Here we use $A_0$ for the total relative weight of the $N_p$ peak contributions, as opposed to $a_0$ used for the weight with a single $\delta$-function
at the edge. We will denote by $a_0$ also an intermediate peak weight in the modified optimization procedure outlined here.

\begin{figure}
\includegraphics[width=7.5cm]{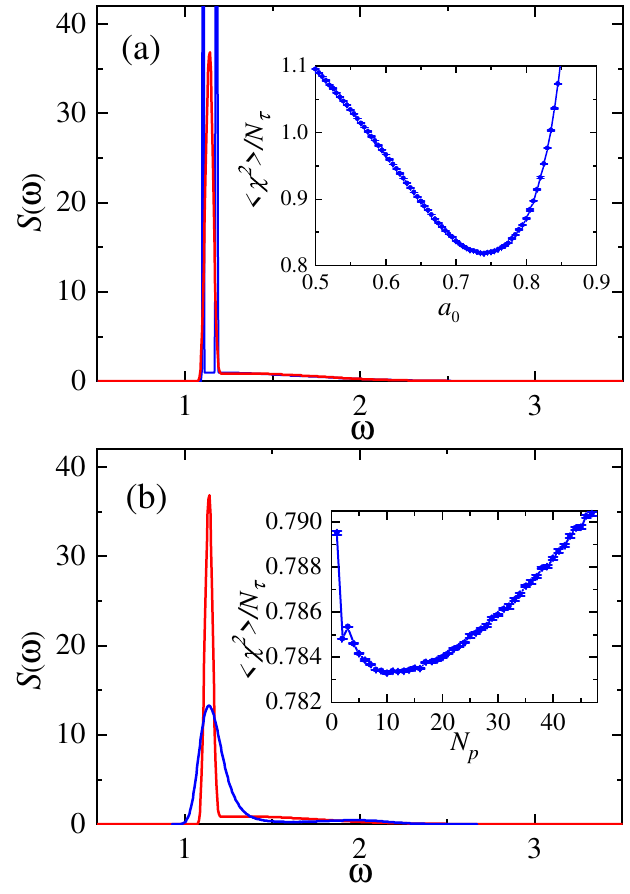}
\caption{Illustration of the two-line scan procedure to determine the optimal $A_0$ and $N_p$ for the $L=256$ chain with $\alpha=2$, $k=3\pi/4$.
The error level of $G(\tau)$ is slightly better than $10^{-5}$. In all the cases, $N_c=2000$ and the sampling temperature is $\Theta=0.1$. The initial step with
$N_p=2$ to fix $a_0$ is illustrated in (a), with the goodness-of-fit scan shown in the inset. The total peak weight $A_0$ for the second scan is taken as the optimal
$a_0=0.74$ plus the continuum weight from the lower peak to the mid point between the two spikes; here $p_0=0.014$. The optimal $N_p=2$ spectrum (blue) is compared
with the final result after $N_p$ has been fixed in the second scan (red). The final spectrum in (b), shown in red, is produced with the optimal $(A_0,N_p)=(0.754,10)$,
with the goodness-of-fit scan over $N_p$ with $A_0$ fixed shown in the inset. The spectrum is is compared with the result of unconstrained SAC (with both amplitude
and frequency updates), shown in blue.}
\label{reala2q097}
\end{figure}

Figure \ref{reala2q097} illustrates the procedure for the $L=256$ chain with $\alpha=2.0$ at $k=3\pi/4$. In the first scan to determine a (near) optimal
weight $a_0$ (which is not exactly the same as $A_0$, as explained below), the number of peak $\delta$-functions is set to $N_p=2$. The idea here is that these two
contributions can bracket a narrow peak and roughly account for its width and weight. The constraint of the continuum $\delta$-functions at this stage is that they
can extend all the way to the lower of the two peak $\delta$-functions, instead of being only above the mean location of the peak as in Fig.~\ref{Fig.sacpara3}.
The entropic downward pressure exerted by the continuum on the lower edge will then lead to a more optimal bracketing of a broadened peak. Here it should be noted
that there can also be significant spectral weight from the continuum contributions between the two peak $\delta$-functions, and to account for this weight
in the second scan, where there will be no continuum contributions below the mean peak location, we take the total peak weight as $A_0=a_0+p_0$, where $p_0$
is the weight between the low edge and the mean $2$-$\delta$ location.

Figure \ref{reala2q097}(a) shows the optimal $2$-$\delta$ peaked spectrum with $a_0=0.740$ (the optimal value according to the scan shown in the inset of the
figure) and $A_0=a_0+p_0=0.754$. We have also included in Fig.~\ref{reala2q097}(a) the final spectrum after the second scan, to show how the two peaks clearly
bracket the final broadened peak. The continua above the peaks are very similar in these two spectra. The second scan, further illustrated in
Fig.~\ref{reala2q097}(b), is over $N_p$ with $A_0$ fixed. The number of $\delta$-functions in the continuum is $N_c=2000$ and the optimal fraction of the peak
contributions is $N_p/N_c=0.005$. The small value of this ratio is why there is much less entropic broadening of the multi-$\delta$ peak than the continuum, thus
allowing a narrow peak to form according to its dominant impact on $\chi^2$ in the sampling. The final result is shown in Fig.~\ref{reala2q097}(b) along
with the result of unconstrained sampling. Here it can be noted that the continuum above the narrow peak does not have any additional peak, unlike the unconstrained
case. The second peak there is likely spurious, being a consequence of too much weight above the first peak, which necessitates a compensating effect of reduced
weight, leading to a minimum, at slightly higher energy.

\subsection{Resolution check}\label{appsubsec:resolution}

The multi-$\delta$-peak parametrization can resolve significantly sharp peaks than the conventional unconstrained SAC method, but what is the limit of the sharpness
detectable by SAC? As discussed in Ref.~\onlinecite{Shao23}, we can perform a resolution check for the $G(\tau)$ data we are working with to determine whether the
extracted width of the leading peak is within the resolution of the method itself or not. The resolution check is performed by replacing the peak part of the extracted
spectrum by a macroscopic $\delta$ function whose position is at the mean of the peak part and whose weight is the same as the total peak weight $A_0$. This replacement
corresponds to a modified correlation function $G_\delta(\tau)$, to which noise is also added in the way discussed in Ref.~\onlinecite{Shao23}. The so obtained new
$G_\delta(\tau)$ and covariance matrix (where only the diagonal terms are affected by the added noise, in the absence of more detailed information of how the
peak replacement affects the covariance matrix). A scan over $N_p$ is again performed to find its new optimum. If the new extracted spectrum $S_\delta(\omega)$
at the new optimal $N_p$ is sharper than the original spectrum $S(\omega)$ from $G(\tau)$, then we deem the width of the original peak reliable (the spectrum passes
the resolution check), while an unchanged or wider new peak suggests that the true peak is narrower and cannot be resolved with the present $G(\tau)$ data quality
(the spectrum fails the resolution test).

\begin{figure}
\includegraphics[width=7.5cm]{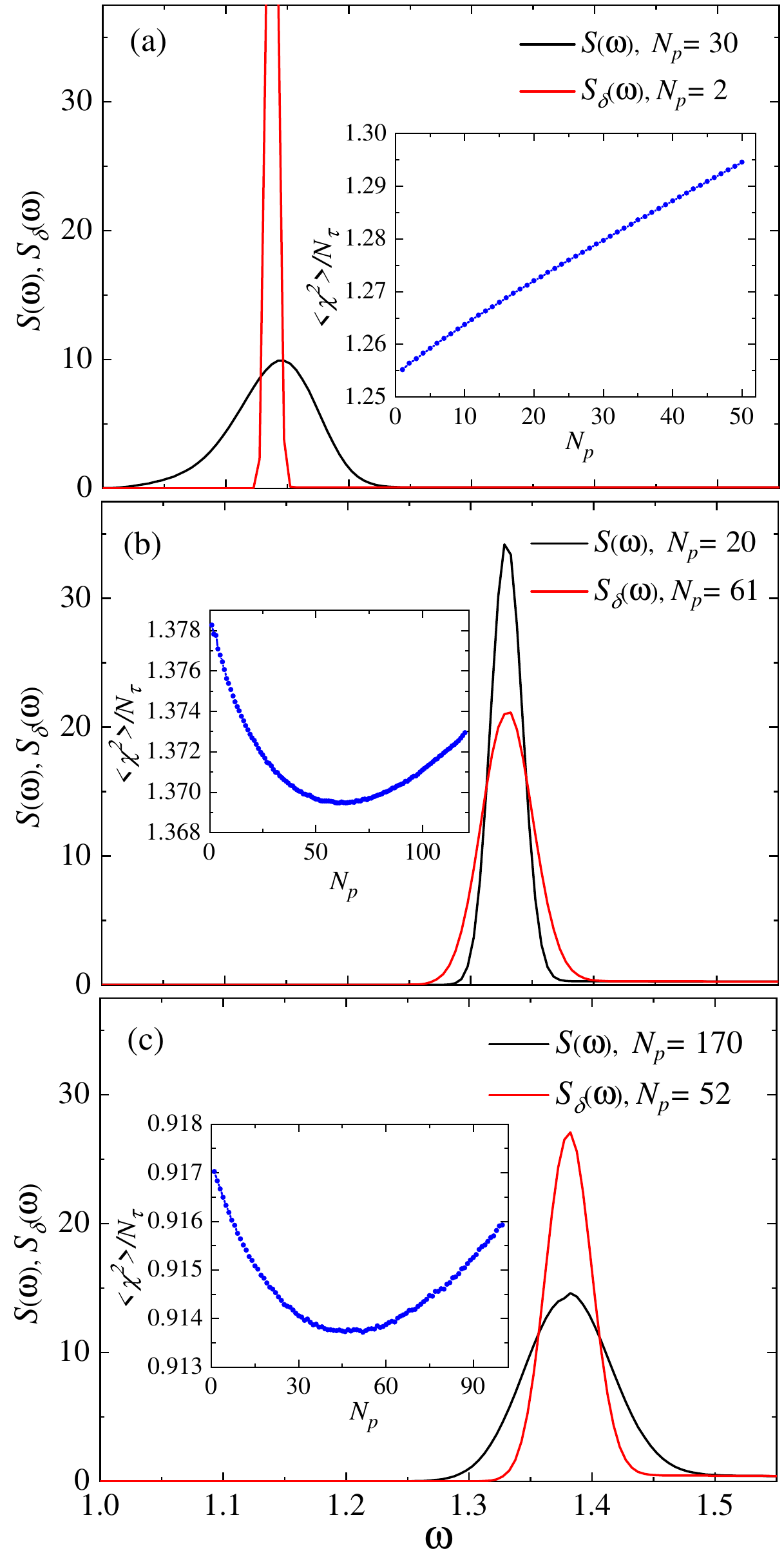}
\caption{Resolution checks for three spectra used in Fig.~\ref{width_combine} at $\alpha=2$, with $k=\pi/4$ in (a), $49\pi/128$ in (b), and $59\pi/128$ in (c), where
(b) corresponds to the region where a minimum width is seen in Fig.~\ref{width_combine}(b), while (a) and (c) correspond to spectra with significantly larger width
to the left and right of the minimum. The insets show goodness-of-fit scans over $N_p$ when the contributions from broadened peak to $G(\tau)$ has been replaced by
that of a single $\delta$-function with the same weight. In each case, the original spectrum $S(\omega)$ is shown with the black curve and the red curve shows the
spectrum at $S_\delta$ sampled at the new optimal $N_p$ value. In (a) and (c), the new spectrum has a taller peak, thus indicating that the original peak is properly
resolvable, while in (b) the new peak is broader, indicating a failed resolution test, i.e., the actual spectrum has a sharper peak than the original one
produced by SAC.}
\label{a2resotest}
\end{figure}

Figure~\ref{a2resotest} shows the results of resolution checks for three spectra obtained at $\alpha=2$ and $k$ values corresponding to the regions of the
two maxima on either side of the width minimum in \ref{width_combine}. Here we can see that the cases $k=\pi/4$ and $59\pi/128$, shown before and after the peak
replacement in Figs.~\ref{a2resotest}(a) and Fig.~\ref{a2resotest}(c), respectively, pass the resolution check, while the $k=49\pi/128$ spectrum in
Fig.~\ref{a2resotest}(b) fails the test. Thus, the minimum width is resolution limited.

While this resolution check is compelling, as we have discussed at length in the main paper, our conclusion based on other parametrizations is that the multi-$\delta$
parametrization is not appropriate for the long-range interacting model. Nevertheless, the parametrization and resolution tests still have some utility in
demonstrating an interesting $k$ dependence of the width of the spectrum even though we have not determined the full spectral profiles accurately.

\section{Self-adjusted edge shape}
\label{app:power_adjust}

\begin{figure}
  \centering
  \includegraphics[width=7.5cm]{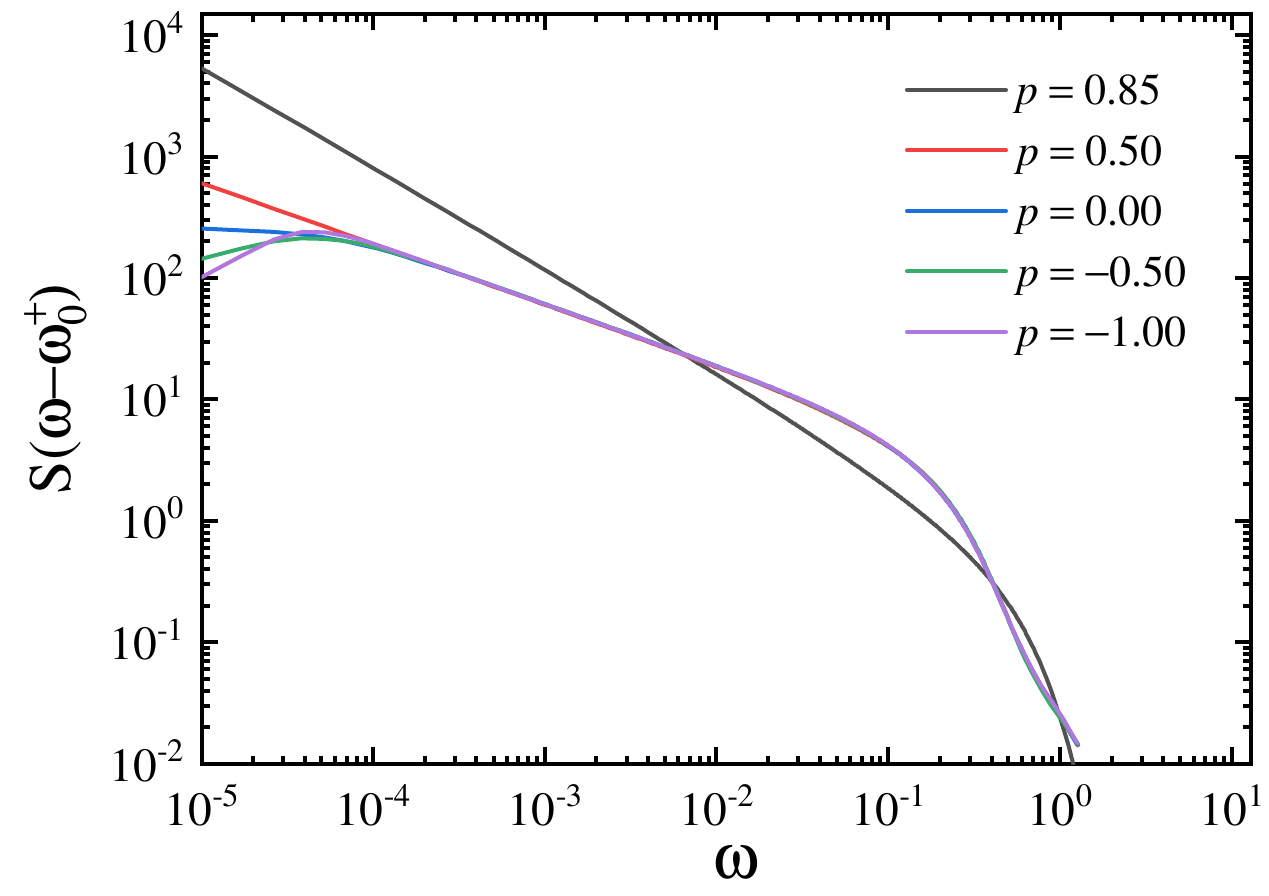}
  \caption{Resultant dynamic structure factors $S(\omega-\omega^+_0)\propto(\omega-\omega^+_0)^{-p}$ with different initial setup of the power $p$ from
    Fig.~\ref{Fig.JQ3LR_SAC_05pi}(e) using edge SAC.}\label{Fig.power_adjust}
\end{figure}

In Sec.~\ref{subsec:afmocase}, we applied SAC with a tunable edge exponent $p$ to produce the power-law spectrum with optimal $p=0.85$, shown in
Fig.~\ref{Fig.JQ3LR_SAC_05pi}(e). It is interesting to observe in the inset of Fig.~\ref{Fig.JQ3LR_SAC_05pi}(e) that $\langle \chi^2\rangle$ is essentially constant
for $p<0.5$, indicating that a similarity of the resultant spectra within a large range of the $p$ values that, at face value, seem to impose very different edge
shapes. The reason for this behavior is illustrated in Fig.~\ref{Fig.power_adjust}, which shows results for $S(\omega-\omega_0)$ obtained with five different $p$
values. Here the spectra for $p \le 0.5$ almost fall on top of each other, except for the very small region $\omega-\omega_0 \lesssim 10^{-4}$. At higher energies,
the $p < 0.5$ curves coincide almost exactly with that for $p=0.5$. The rapid cross-over to the $p=0.5$ form takes place because the sampled cross-over location
$i_c$ in index space (of the $\delta$-functions with $\omega_i$ increasing with $i$) are pushed to small values in order for the $G(\tau)$ fits to be good. In
contrast, for $p=0.85$ the cross-over is not observed clearly because it takes place at large (and fluctuating) $i_c$ and the shape of the spectrum overall is
different from that at $p \le 0.5$. The  $\langle \chi^2\rangle$ minimum in Fig.~\ref{Fig.JQ3LR_SAC_05pi}(e) identifies $p\approx 0.85$ as the optimal
exponent in this case.

\bibliographystyle{apsrev4-2}

\end{document}